\documentclass[11pt]{llncs}
\usepackage{amssymb,amsmath,enumerate,graphicx}
\usepackage{verbatim}
\usepackage{cite}
\usepackage[usenames]{color}

\usepackage{tikz}
\usetikzlibrary{snakes}
\tikzstyle{tre}=[circle,draw,minimum size=3mm]
\tikzstyle{btre}=[circle,draw,minimum size=4mm]
\newcommand{\etq}[1]{%
\draw (#1) node {\scriptsize $#1$};}

\newcommand{\pathgr}{\!\rightsquigarrow\!{}}

\renewcommand{\leq}{\leqslant}
\renewcommand{\geq}{\geqslant}
\renewcommand{\le}{\leqslant}
\renewcommand{\ge}{\geqslant}
\newcommand{\NN}{\mathbb{N}}
\newcommand{\RR}{\mathbb{R}}
\newcommand{\RRz}{\mathbb{R}_{\geq 0}}
\newcommand{\RRp}{\mathbb{R}_{>0}}
\newcommand{\TT}{\mathcal{T}}
\newcommand{\bTT}{\mathcal{BT}}
\newcommand{\nTT}{\mathcal{NT}}

\begin{document}

\title{Nodal distances for rooted phylogenetic trees}

\author{Gabriel Cardona\inst{1} \and Merc\`e Llabr\'es\inst{1}\inst{2} \and Francesc Rossell\'o\inst{1}\inst{2} \and
Gabriel Valiente\inst{2}\inst{3}} \authorrunning{G. Cardona et al.}
\institute{Department of Mathematics and Computer Science, University
of the Balearic Islands, E-07122 Palma de Mallorca, Spain
\and
Research Institute of Health Science (IUNICS),  E-07122 Palma de Mallorca, Spain
\and
Algorithms, Bioinformatics, Complexity and Formal Methods Research
Group, Technical University of Catalonia, E-08034 Barcelona, Spain}

\maketitle

\begin{abstract}
Dissimilarity measures for (possibly weighted) phylogenetic
trees based on the comparison of their vectors of path lengths between
pairs of taxa, have been present in the systematics
literature since the early seventies.  But, as far as rooted phylogenetic trees goes, 
 these vectors can only
separate non-weighted binary trees, and therefore
these dissimilarity measures are metrics only on this class.  In this paper we overcome this problem, by
splitting in a suitable way each path length between two taxa into two
lengths.  We prove that the resulting \emph{splitted path lengths
matrices} single out arbitrary rooted phylogenetic trees with nested
taxa and arcs weighted in the set of positive real numbers.  This
allows the definition of metrics on this general class  by comparing these matrices by means of metrics in
spaces $\mathcal{M}_n(\RR)$ of real-valued $n\times n$ matrices.  We conclude this
paper by establishing some basic facts about the metrics for
non-weighted phylogenetic trees defined in this way using $L^p$ metrics on
$\mathcal{M}_n(\RR)$, with $p\in\NN\setminus\{0\}$.
\end{abstract}

\section{Introduction}
\label{sec:intro}

The exponential increase in the amount of available genomic and
metagenomic data has produced an explosion in the number of
phylogenetic trees proposed by researchers: according to Rokas
\cite{rokas:06}, phylogeneticists are currently publishing an average
of 15 phylogenetic trees per day.  Many such trees are alternative
phylogenies for the same sets of organisms, obtained from different
datasets or using different evolutionary models or different phylogenetic reconstruction algorithms  \cite{hoef:05}.  This
variety of phylogenetic trees makes it
necessary the existence of methods for measuring the differences
between phylogenetic trees \cite[Ch.~30]{fel:04}, and the safest way to quantify these differences 
is by using a metric, for which zero difference means isomorphism.

The comparison of phylogenetic trees is also used to assess the
stability of reconstruction methods \cite{willcliff:taxon71}, and it
is essential to performing phylogenetic queries on databases~\cite{page:2005}.  Further, the need for comparing
phylogenetic trees also arises in the comparative analysis of
clustering results obtained using different methods or
different distance matrices, and there is a growing interest in the
assessment of clustering results in
bioinformatics~\cite{handl.ea:2005}.  Recent applications of
the comparison of phylogenetic-like trees also include the study of the similarity
between sequences, or sets of sequences, by measuring the difference
between their context trees \cite{leonardi.ea:08}.
In summary, and using the words of Steel and Penny~\cite{steelpenny:sb93}, tree comparison metrics are an important aid in the study of evolution.

 Many metrics for phylogenetic tree
comparison have been proposed so far, including the Robinson-Foulds,
or partition, metric~\cite{robinson.foulds:1979,robinson.foulds:mb81}, the
nearest-neighbor interchange metric~\cite{waterman.smith:1978}, the
subtree transfer distance~\cite{allen.steel:2001}, and the triples
metric \cite{critchlow.ea:1996}.
In the early seventies, several researchers proposed
dissimilarity measures for  (possibly weighted) rooted phylogenetic trees based
on the comparison of the vectors of lengths of paths connecting pairs of
taxa. The aim of these measures was to quantify the rate at which pairs of taxa that are close together in one tree lie at opposite ends in another tree~\cite{pennyhendy:sz85}. These authors defined the dissimilarity between a pair of trees 
as the euclidean distance between the
corresponding vectors of path lengths \cite{farris:sz69,farris:sz73},
the Manhattan distance between these vectors \cite{willcliff:taxon71}
or the correlation between these vectors \cite{phipps:sz71}.  Similar
dissimilarity measures have also been defined for unrooted
phylogenetic trees \cite{bluis.ea:2003,steelpenny:sb93}. Although
different names have been used for these dissimilarity measures
(cladistic difference \cite{farris:sz69}, topological distance
\cite{phipps:sz71}, path difference distance \cite{steelpenny:sb93}),
the term \emph{nodal distance} seems to have prevailed
\cite{bluis.ea:2003,puigbo.ea:07}.
According to Steel and Penny~\cite{steelpenny:sb93}, they have several interesting features that make them deserve more study and consideration.

The theoretical basis for these nodal distances is Smolenskii's
theorem \cite{smolenskii:63} establishing that two unrooted
phylogenetic trees $T,T'$ on the same set $S$ of taxa are isomorphic
if, and only if, for every pair of leaves $i,j$, the distances between
$i$ and $j$ in $T$ and in $T'$ are the same.  This result was later
expanded by Zaretskii \cite{zaretskii:1965}, who characterized the
vectors of distances between pairs of leaves of an unrooted
phylogenetic tree by means of the well-known four-point condition.
Smolenskii's and Zaretskii's papers were published in
Russian, and it has contributed to the fact that their results have been rediscovered and
generalized many times~\cite{simoes.ea:1990,boesch:78,buneman:71,simoes:69}; for a
modern textbook treatment of these results in all their generality (weighted unrooted trees with nested taxa),
see \cite[Ch.~7]{semple.steel:03}, and for a historical account, see
\cite{abdi:90}.

Unfortunately, Smolenskii's theorem is not valid for arbitrary 
rooted phylogenetic trees: there exist  non-isomorphic  rooted
phylogenetic trees with the same path lengths between pairs of leaves
(see Figs.~\ref{fig:contr1}, \ref{fig:element}, \ref{fig:contr2}).  It turns out that only the \emph{fully
resolved}, or \emph{binary}, \emph{non-weighted} rooted phylogenetic trees are singled out by
their path lengths vectors, and therefore the nodal distances based on the comparison of these vectors are metrics (more
specifically, zero nodal distance means isomorphism) only on the space
of non-weighted binary phylogenetic trees. Although this result seems to be known
since the time of the first proposals of nodal distances, we have not
been able to find an explicit proof in the literature, and thus, for
the sake of completeness, we include a simple proof of this fact in Section \ref{sec:bintrees}, reducing it
to the general version of Smolenksii's result.

The main result of this paper is the definition of metrics on the space of \emph{arbitrary} rooted phylogenetic trees  that generalize the nodal distances, where arbitrary means non necessarily binary and with possibly nested taxa and arcs weighted in the set  of positive real numbers. To do that, we split
each path between two taxa into the paths from their
least common ancestor to each taxa.  In this way we associate to each rooted
phylogenetic tree with $n$ taxa an $n\times n$ matrix, with rows and
columns indexed by the taxa, whose $(i,j)$-entry contains the length of
the path from the least common ancestor of the $i$-th and $j$-th taxa
to the $i$-th taxon.  We prove that these \emph{splitted path lengths
matrices} single out arbitrary rooted phylogenetic trees, and then we 
use them to define \emph{splitted nodal metrics} on the space of weighted
rooted phylogenetic trees with nested taxa by comparing these matrices through 
real-valued norms applied to their difference.
We also prove some basic properties of the splitted nodal metrics on the
space of non-weighted rooted phylogenetic trees
obtained using the $L^p$ norms, with $p\in \NN\setminus\{0\}$.

\section{Notations and conventions}
\label{sec:prel}

A \emph{rooted tree} is a non-empty directed finite graph that contains a
distinguished node, called the \emph{root}, from which every other
node can be reached through exactly one path.  An \emph{$A$-weighted
rooted tree}, with $A\subseteq \RR$, is a pair $(T, \omega)$
consisting of a rooted tree $T=(V,E)$ and a \emph{weight function}
$\omega: E\to A$ that associates to every arc $e\in E$ a real number
$\omega(e)\in A$.  In this paper we shall only consider two sets $A$
of weights: the set of non-negative real numbers $\RRz=\{t\in \RR\mid
t\geq 0\}$, and the set of positive real numbers $\RRp=\{t\in \RR\mid
t> 0\}$.  When the set $A$ is irrelevant (for instance, in general
definitions), we shall omit it and simply talk about \emph{weighted},
instead of $A$-weighted, trees.  We identify every \emph{non-weighted}
(that is, where no weight function has been explicitly defined) rooted
tree $T$ with the weighted rooted tree $(T,\omega)$ with $\omega$ the
weight 1 constant function.

Let $T=(V,E)$ be a rooted tree.  Whenever $(u,v)\in E$, we say that
$v$ is a \emph{child} of $u$ and that $u$ is the \emph{parent} of $v$.
Every node in $T$ has exactly one parent, except the root, which has
no parent.  The number of children of a node is its \emph{out-degree}.
The nodes without children are the \emph{leaves} of the tree, and the
other nodes are called \emph{internal}.  An arc $(u,v)$ is
\emph{internal} when its head $v$ is internal, and \emph{pendant} when
$v$ is a leaf.  The out-degree 1 nodes are called \emph{elementary}.
A tree is \emph{binary} when all its internal nodes have out-degree 2.

Given a path $(v_{0},v_{1},\ldots,v_{k})$ in a rooted tree $T$, its
\emph{origin} is $v_{0}$, its \emph{end} is $v_{k}$, and its
\emph{intermediate nodes} are $v_{1},\ldots,v_{k-1}$.  Such a path is
\emph{non-trivial} when $k\geq 1$.  We shall represent a path
\emph{from $u$ to $v$}, that is, a path with origin $u$ and end $v$,
by $u \pathgr v$.  Whenever there exists a (non-trivial) path
$u\pathgr v$, we shall say that $v$ is a (\emph{non-trivial})
\emph{descendant} of $u$ and also that $u$ is a (\emph{non-trivial})
\emph{ancestor} of $v$.  If $v$ is a descendant of $u$, the path
$u\pathgr v$ is unique.  The \emph{distance} from a node $u$ to a
descendant $v$ of it in a weighted rooted tree is the sum of the
weights of the arcs forming the unique path $u\pathgr v$; in a
non-weighted rooted tree, this distance is simply the number of arcs
of this path. The \emph{depth}  of a node $v$,  in symbols $\mathrm{depth}_T(v)$, is the distance from the root to $v$.

The \emph{least common ancestor} (LCA) of a pair of nodes
$u,v$ of a rooted tree $T$, in symbols
$[u,v]_T$, is the unique common ancestor of them that is
a descendant of every other common ancestor of them.  Alternatively,
it is the unique common ancestor of $u,v$ such that the
paths from it to $u$ and $v$ have only their origin in common.  In
particular, if  one of the nodes, say $u$, is an ancestor of
the other, then $[u,v]_T=u$.

Let $S$ be a non-empty finite set of  \emph{labels}, or \emph{taxa}.  A
(\emph{weighted}) \emph{phylogenetic tree} on $S$ is a (weighted)
rooted tree with some of its nodes, including all its leaves and its
elementary nodes, bijectively labeled in the set $S$.  In such a phylogenetic
tree, we shall always identify, usually without any further mention, a
labeled node with its taxon. The internal
labeled nodes of a phylogenetic tree are called \emph{nested taxa}.

Two phylogenetic trees $T$ and $T'$ on the same set $S$ of taxa are
\emph{isomorphic} when they are isomorphic as directed graphs and the
isomorphism sends each labeled node of $T$ to the labeled node with the same label in $T'$; an isomorphism of weighted
phylogenetic trees is also required to preserve arc weights.  As
usual, we shall use the symbol $\cong$ to denote the existence of an
isomorphism.

Although our main object of study are the weighted phylogenetic trees,
and hence they are rooted trees, in the next section there will also
appear unrooted trees.  An \emph{unrooted tree} is an undirected
finite graph where every pair of nodes is connected by exactly one
path.  An \emph{$A$-weighted unrooted tree} is a pair $(T, \omega)$
consisting of an unrooted tree $T=(V,E)$ and a \emph{weight function}
$\omega: E\to A$.  The \emph{distance} between two nodes in a weighted
unrooted tree is the sum of the weights of the edges forming the
unique path that connects these nodes.

An unrooted tree is \emph{partially labeled} in a set $S$ when some of
its nodes are bijectively labeled in the set $S$.  An \emph{unrooted
$S$-tree} is an unrooted tree partially labeled in $S$ with all its
leaves and all its nodes of degree 2 labeled.

Given a phylogenetic tree $T=(V,E)$ on $S$, its \emph{unrooted
version} is the unrooted tree $T^u=(V,E^u) $ partially labeled in $S$
obtained by replacing each arc $(u,v)\in E$ by an edge $\{u,v\}\in
E^u$, and keeping the labels.

The notion of isomorphism for (possibly weighted) partially labeled
unrooted trees   is similar to the notion given
in the rooted case.  Notice that if $T_1=(V_1,E_1)$ and $T_2=(V_2,E_2)$ are two
phylogenetic trees on the same set $S$ of taxa, with roots $r_1$ and
$r_2$, respectively, then a mapping $f:V_1\to V_2$ is an isomorphism
between $T_1$ and $T_2$ if, and only if, it is an isomorphism between
$T_1^u$ and $T_2^u$ and $f(r_1)=r_2$.

\section{Path lengths separate non-weighted binary phylogenetic trees}
\label{sec:bintrees}

Let $T$ be an $\RRz$-weighted phylogenetic tree on the set
$S=\{1,\ldots,n\}$.  For every $i,j\in S$, let $\ell_T(i,j)$ and
$\ell_T(j,i)$ denote the distances from $[i,j]_T$ to $i$ and $j$,
respectively.  The \emph{path length} between two labeled nodes $i$
and $j$ is
$$
L_T(i,j)=\ell_T(i,j)+\ell_T(j,i).
$$

\begin{definition}
The \emph{path lengths  vector} of $T$ is  the  vector 
$$
L(T)=\big(L_T(i,j)\big)_{1\leq i<j\leq n}\in \RR^{n(n-1)/2},
$$
with its entries ordered lexicographically in $(i,j)$.
\end{definition}

These path length vectors have been used since the early seventies to
compare non-weighted, binary phylogenetic trees
\cite{farris:sz69,phipps:sz71,willcliff:taxon71}, but we have not been
able to find an explicit proof in the literature of the fact that this
kind of phylogenetic trees can be singled out by means of their path
lengths vector.  For the sake of completeness, we provide here a
simple proof of this fact, derived from Smolenskii's theorem
\cite{smolenskii:63} that establishes that the vector of distances
between pairs of labeled nodes characterizes up to isomorphism an
$\RRp$-weighted unrooted $S$-tree; see also Thm.~7.1.8 in
\cite{semple.steel:03}.

\begin{proposition}
\label{prop:nod-bintree}
Two non-weighted binary phylogenetic trees on the same set $S$ of taxa are
isomorphic if, and only if, they have the same path lengths vectors.
\end{proposition}

\begin{proof}
The `only if' implication is obvious.  As far as the `if' implication
goes, let $T_1$ and $T_2$ be two non-weighted binary phylogenetic
trees on the same set $S$ with the same path lengths vectors.  If
$|S|= 1$, the equivalence in the statement is obvious, because every phylogenetic tree with only one labeled node consists only of one node. So we assume henceforth that $|S|\geq 2$.

For every $t=1,2$, let $(T_t^*,\omega_t)$ be the $\RRp$-weighted unrooted
$S$-tree defined as follows:
\begin{itemize}
\item If the root of $T_t$ is labeled, then $T_t^*=T_t^u$ and all
edges of $T_t^*$ have weight 1.

\item If the root $r_t$ of $T_t$ is not labeled, and if $u_t,v_t$ are
the children of $r_t$, then $T_t^*$ is obtained from $T_t^u$ by
removing the node $r_t$ and replacing the edges
$\{r_t,u_t\},\{r_t,v_t\}$ by a single edge $\{u_t,v_t\}$, and then all
edges of $T_t^*$ have weight 1, except $\{u_t,v_t\}$, which has weight
2.
\end{itemize}
It is straightforward to check that such a $T_{t}^{*}$ is always an
unrooted $S$-tree: the root $r_{t}$ of $T_{t}$ is the only degree 2
node in $T_{t}^u$ and then, if it is labeled, $T_{t}^u$ is an unrooted
$S$-tree, and if it is non labeled, we remove it in the construction
of $T_{t}^{*}$ without modifying the degrees of the remaining nodes.  Moreover, it is also obvious from the construction
that the distance between any pair of labeled nodes in $T_{t}^{*}$ is
equal to the path length between these nodes in $T_{t}$.  In
particular, $T_1^*$ and $T_2^*$ have the same distances between each
pair of labeled nodes.  Then, by
\cite[Thm.~7.1.8]{semple.steel:03}.

$T_1^*\cong T_2^*$ as weighted unrooted $S$-trees. 

It remains to check that this isomorphism induces an isomorphism of
phylogenetic trees $T_1\cong T_2$.  
To do it, notice that, since the isomorphism between $T_1^*$ and $T_2^*$ preserves edge weights,  there are only two possibilities:
\begin{itemize}
\item All edges in $T_1^*$ and $T_2^*$ have weight 1. In this case
$T_1^*=T_1^u$ and $T_2^*=T_2^u$ and the isomorphism $T_1^u\cong T_2^u$ sends the root of $T_1$ to the root of $T_2$, because they are the only degree 2 nodes in $T_1^*$ and $T_2^*$.
Therefore, it induces an isomorphism $T_1\cong T_2$. 

\item Both $T_1^*$ and $T_2^*$ have one weight 2 edge, say $\{u_1,v_1\}$ and $\{u_2,v_2\}$, respectively. Then each
$T_t^u$ is obtained from $T_t^*$ by adding the root $r_t$ of $T_t$ and
splitting the edge $\{u_t,v_t\}$ into two edges
$\{u_t,r_t\}$ and $\{v_t,r_t\}$.  Since the isomorphism $T_1^*\cong
T_2^*$  sends $\{u_1,v_1\}$ to
$\{u_2,v_2\}$, its extension to a mapping $V_1\to V_2$ by
sending $r_1$ to $r_2$ defines an isomorphism $T_1^u\cong T_2^u$ that
sends the root of $T_1$ to the root of $T_2$, and hence an isomorphism
$T_1\cong T_2$.  \qed
\end{itemize}
\end{proof}

Let $\bTT_n$ be the class of all non-weighted binary phylogenetic
trees on $S=\{1,\ldots,n\}$.  The injectivity up to isomorphisms of the mapping
$$
\begin{array}{rcl}
L: \bTT_n & \to & \RR^{n(n-1)/2}\\
 T & \mapsto & L(T)
 \end{array}
 $$ 
makes the classical definitions of \emph{nodal metrics} on
$\bTT_n$ induced by metrics on $\RR^{n(n-1)/2}$ to yield, indeed, metrics.  For example, recall
that the \emph{$L^p$ norm} on $\RR^m$ is defined as
$$
\|(x_1,\ldots,x_m)\|_p=\left\{
\begin{array}{ll}
\big|\{i\mid i=1,\ldots,m,\  x_i\neq 0\}\big| & \mbox{if $p=0$}\\
\sqrt[p]{\sum_{i=1}^m |x_i|^p} & \mbox{if $p\in \NN^+$}\\
\max\{|x_i|\mid i=1,\ldots,m\} & \mbox{if $p=\infty$}
\end{array}
\right.
$$
where, here and henceforth, $\NN^+$ stands for $\NN\setminus\{0\}$.  Each $L^p$ norm on
$\RR^{n(n-1)/2}$ induces then a metric on $\bTT_n$ through the formula
$$
d_p(T_1,T_2)=\|L(T_1)-L(T_2)\|_p.
$$
Some of these metrics have been present in the literature since the
early seventies.  For instance, Farris \cite{farris:sz69} introduced
the metric on $\bTT_n$ induced by the $L^2$, or Euclidean, norm on
$\RR^{n(n-1)/2}$:
$$
d_2(T_1,T_2)=\sqrt{\sum_{1\leq i<j\leq n}(L_{T_1}(i,j)-L_{T_2}(i,j))^2}
$$
(he called it \emph{cladistic difference}), while Williams and
Clifford \cite{willcliff:taxon71} proposed the metric on $\bTT_n$
induced by the $L^1$, or Manhattan, norm on $\RR^{n(n-1)/2}$:
$$
d_1(T_1,T_2)=\sum_{1\leq i<j\leq n}|L_{T_1}(i,j)-L_{T_2}(i,j)|.
$$

\begin{figure}[htb]
\begin{center}
            \begin{tikzpicture}[thick,>=stealth,scale=0.4]
              \draw(0,0) node[tre] (1) {}; \etq 1
              \draw(2,0) node[tre] (2) {}; \etq 2
              \draw(4,0) node[tre] (3) {}; \etq 3
              \draw(6,0) node[tre] (4) {}; \etq 4
              \draw(5,2) node[tre] (a) {}; %\etq 3  
\draw(1,4) node[tre] (r) {}; %\etq 3      
            \draw[->] (r)--(1);
            \draw[->] (r)--(2);
                        \draw[->] (r)--(a);
            \draw[->] (a)--(3);
            \draw[->] (a)--(4);
              \draw(3,-1) node  {$T$};
            \end{tikzpicture}
  \qquad\qquad
%  \begin{tikzpicture}[thick,>=stealth,scale=0.4]
%            \draw(0,0) node[tre] (3) {}; \etq 3
%              \draw(2,0) node[tre] (4) {}; \etq 4
%              \draw(4,0) node[tre] (1) {}; \etq 1
%              \draw(6,0) node[tre] (2) {}; \etq 2
%              \draw(5,2) node[tre] (a) {}; %\etq 3  
%\draw(1,4) node[tre] (r) {}; %\etq 3      
%            \draw[->] (r)--(3);
%            \draw[->] (r)--(4);
%                        \draw[->] (r)--(a);
%            \draw[->] (a)--(1);
%            \draw[->] (a)--(2);
%              \draw(3,-1) node  {$T'$};
%  \end{tikzpicture}
            \begin{tikzpicture}[thick,>=stealth,scale=0.4]
              \draw(0,0) node[tre] (1) {}; \etq 1
              \draw(2,0) node[tre] (2) {}; \etq 2
              \draw(4,0) node[tre] (3) {}; \etq 3
              \draw(6,0) node[tre] (4) {}; \etq 4
              \draw(5,4) node[tre] (a) {}; %\etq 3  
\draw(1,2) node[tre] (r) {}; %\etq 3      
            \draw[->] (r)--(1);
            \draw[->] (r)--(2);
                        \draw[->] (a)--(r);
            \draw[->] (a)--(3);
            \draw[->] (a)--(4);
              \draw(3,-1) node  {$T'$};
            \end{tikzpicture}
\end{center}
\caption{\label{fig:contr1} 
Two non-isomorphic non-binary phylogenetic trees with the same path
lengths vectors.}
\end{figure}
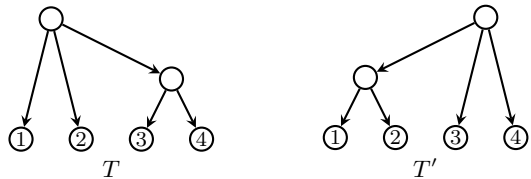

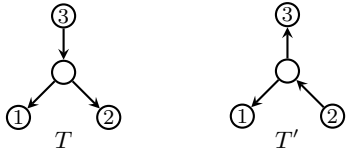
\begin{figure}[htb]
\begin{center}
            \begin{tikzpicture}[thick,>=stealth,scale=0.3]
              \draw(0,0) node[tre] (1) {}; \etq 1
              \draw(4,0) node[tre] (2) {}; \etq 2
              \draw(2,2) node[tre] (a) {};  
              \draw(2,4.5) node[tre] (3) {}; \etq 3
             \draw[->] (a)--(1);
            \draw[->] (a)--(2);
                        \draw[->] (3)--(a);
              \draw(2,-1) node  {$T$};
             \end{tikzpicture}
  \qquad\qquad
%  \begin{tikzpicture}[thick,>=stealth,scale=0.3]
%               \draw(0,0) node[tre] (1) {}; \etq 1
%              \draw(4,0) node[tre] (3) {}; \etq 3
%              \draw(2,2) node[tre] (a) {};  
%              \draw(2,4.5) node[tre] (2) {}; \etq 2
%             \draw[->] (a)--(1);
%            \draw[->] (a)--(3);
%                        \draw[->] (2)--(a);
%              \draw(2,-1) node  {$T'$};
%  \end{tikzpicture}
            \begin{tikzpicture}[thick,>=stealth,scale=0.3]
              \draw(0,0) node[tre] (1) {}; \etq 1
              \draw(4,0) node[tre] (2) {}; \etq 2
              \draw(2,2) node[tre] (a) {};  
              \draw(2,4.5) node[tre] (3) {}; \etq 3
             \draw[->] (a)--(1);
            \draw[->] (2)--(a);
                        \draw[->] (a)--(3);
              \draw(2,-1) node  {$T'$};
             \end{tikzpicture}
\end{center}
\caption{\label{fig:element} 
Two non-isomorphic phylogenetic trees with an elementary node and the
same path lengths vectors.}
\end{figure}

Unfortunately, the path lengths vectors cannot be used to separate
phylogenetic trees in much more general classes than the one considered in the
previous proposition. For instance, they does not single out
phylogenetic trees with nodes of out-degree greater than 2 (see
Fig.~\ref{fig:contr1}),   phylogenetic trees with (labeled) elementary
nodes (see Fig.~\ref{fig:element}), and  weighted binary
phylogenetic networks with weights different from 1 (see
Fig.~\ref{fig:contr2}).  Therefore, no metric for general phylogenetic
trees can be derived from path lengths  alone. We overcome this problem in the next section.

\begin{figure}[htb]
\begin{center}
            \begin{tikzpicture}[thick,>=stealth,scale=0.3]
              \draw(0,0) node[tre] (1) {}; \etq 1
              \draw(4,0) node[tre] (2) {}; \etq 2
              \draw(2,3) node[tre] (r) {}; %\etq r
             \draw[->] (r)--node[near end, above] {\footnotesize 1} (1);
            \draw[->] (r)--node[near end, above] {\footnotesize 2}  (2);
              \draw(2,-1) node  {$T$};
            \end{tikzpicture}
  \qquad\qquad
  \begin{tikzpicture}[thick,>=stealth,scale=0.3]
              \draw(0,0) node[tre] (1) {}; \etq 1
              \draw(4,0) node[tre] (2) {}; \etq 2
              \draw(2,3) node[tre] (r) {}; %\etq r
             \draw[->] (r)--node[near end, above] {\footnotesize 2} (1);
            \draw[->] (r)--node[near end, above] {\footnotesize 1} (2);
              \draw(2,-1) node  {$T'$};
  \end{tikzpicture}
\end{center}
\caption{\label{fig:contr2} 
Two non-isomorphic $\RRp$-weighted binary phylogenetic trees with the
same path lengths vectors.}
\end{figure}

\begin{remark}
Let $T$ be a non-weighted binary phylogenetic tree on a set $S$ of
taxa.  Since the path lengths vector $L(T)$ is the vector of distances
of a (possibly weighted) unrooted $S$-tree (see the proof of
Proposition \ref{prop:nod-bintree}), it is well-known (see, for
instance, Lem.~{7.1.7} in \cite{semple.steel:03}) that it satisfies
the \emph{four-point condition}: for every $a,b,c,d\in S$,
$$
L_T(a,b)+L_T(c,d)\leq \max\{L_T(a,c)+L_T(b,d),L_T(a,d)+L_T(b,c)\}.
$$
Zaretskii's theorem \cite{zaretskii:1965} establishes that any
dissimilarity measure on $S$ satisfying this four-point condition is
given by the distances between labeled nodes in an $\RRp$-weighted
unrooted $S$-tree (see Thm.~7.2.6 in \cite{semple.steel:03}).  But, to
our knowledge, it is not known what extra properties should be
required to such a dissimilarity measure on $S$ to guarantee that it
is given by the path lengths between labeled nodes in a non-weighted
binary phylogenetic tree.
\end{remark}

\section{Splitted path lengths separate arbitrary phylogenetic trees}
\label{sec:Strees}

Let $(T,\omega)$, with $T=(V,E)$, be again an $\RRz$-weighted
phylogenetic tree on $S=\{1,\ldots,n\}$ and, for every $i,j\in S$, let
$\ell_T(i,j)$ and $\ell_T(j,i)$ still denote the distances from
$[i,j]_T$ to $i$ and $j$, respectively.

\begin{definition}
The \emph{splitted path lengths matrix} of $T$ is the $n\times n$ square
matrix over $\RRz$
$$
\ell(T)=\left(\begin{array}{cccc}
\ell_{T}(1,1)\ & \ell_{T}(1,2)\ &  \ldots\ & \ell_{T}(1,n)\\
\ell_{T}(2,1)\ & \ell_{T}(2,2)\ &   \ldots\ & \ell_{T}(2,n)\\
\vdots &  \vdots &   \ddots & \vdots\\
\ell_{T}(n,1)\ & \ell_{T}(n,2)\ &   \ldots\  & \ell_{T}(n,n)
\end{array}
\right)\in \mathcal{M}_n(\RRz).
$$
\end{definition}
Notice that this matrix need not be symmetrical (see the next
example), but all entries $\ell_T(i,i)$ in its main diagonal are 0.

The splitted path lengths matrix $\ell(T)$ of a tree $T\in \TT_n$ can be computed in optimal $O(n^2)$ time, by 
computing by breadth-first search for each internal node of $T$ the distance to each one of its descendant taxa and the pairs of taxa of which it is the LCA.

\begin{example}
The splitted path lengths matrices of the trees $T$ and $T'$ depicted
in Fig.~\ref{fig:contr1} are
$$
\ell(T)=\left(\begin{array}{llll}
0 & 1 & 1 & 1 \\[-0.5ex]
1 & 0 & 1 & 1 \\[-0.5ex]
2 & 2 & 0 & 1 \\[-0.5ex]
2 & 2 & 1 & 0
\end{array}
\right),\quad
\ell(T')=\left(\begin{array}{llll}
0 & 1 & 2 & 2 \\[-0.5ex]
1 & 0 & 2 & 2 \\[-0.5ex]
1 & 1 & 0 & 1 \\[-0.5ex]
1 & 1 & 1 & 0
\end{array}
\right).
$$

The splitted path lengths matrices of the trees $T$ and $T'$ depicted
in Fig.~\ref{fig:element} are
$$
\ell(T)=\left(\begin{array}{llll}
0 & 1 & 2 \\[-0.5ex]
1 & 0 & 2   \\[-0.5ex]
0 & 0 & 0
\end{array}
\right),\quad
\ell(T')=\left(\begin{array}{llll}
0 & 2 & 1 \\[-0.5ex]
0 & 0 & 0 \\[-0.5ex]
1 & 2& 0 
\end{array}
\right).
$$

The splitted path lengths matrices of the weighted trees $T$ and $T'$
depicted in Fig.~\ref{fig:contr2} are
$$
\ell(T)=\left(\begin{array}{llll}
0 & 1   \\[-0.5ex]
2 & 0  
\end{array}
\right),\quad
\ell(T')=\left(\begin{array}{llll}
0 & 2   \\[-0.5ex]
1 & 0 
\end{array}
\right).
$$
\end{example}

This example shows that the splitted path lengths matrices
can separate pairs of phylogenetic trees that could not be separated
by means of their path lengths vectors.  Our main result in this
section states that these matrices characterize arbitrary
$\RRp$-weighted phylogenetic trees.  To prove it, it is convenient to establish
first some lemmas, and to recall a result from \cite{moulton.ea:08}.

\begin{lemma}
\label{lem:previ0}
Let $T$ be an $\RRp$-weighted phylogenetic tree on $S$.  A label $i\in
S$ is a nested taxon of $T$ if, and only if, $\ell_T(i,j)=0$ for some
$j\neq i$.
\end{lemma}

\begin{proof}
If an internal node of $T$ is labeled with $i$, then taking as $j\in S$ any
descendant leaf of $i$ we have that $[i,j]_T=i$ and hence
$\ell_T(i,j)=0$.  Conversely, if $\ell_T(i,j)=0$, then $[i,j]_T=i$ and
therefore the node $i$ is an ancestor of the node $j$. If $i\neq j$, this can only
happen if $i$ is internal.\qed
\end{proof}

\begin{lemma}
\label{lem:previ}
Let $T$ be an $\RRp$-weighted phylogenetic tree on $S$.  For every
$i\in S$, consider the set of weights
$$
W_{i}=\{\ell_{T}(i,j)\mid j\in S,\ \ell_{T}(i,j)>0\}.
$$
\begin{enumerate}[(a)]

\item $W_{i}=\emptyset$ if, and only if, $i$ is the root of $T$.

\item If $W_{i}\neq \emptyset$, then its smallest element $w_{i}$ is
the weight of the arc with head~$i$.
\end{enumerate}
\end{lemma}

\begin{proof}
As far as far (a) goes, $W_i=\emptyset$ if, and only if,
$\ell_T(i,j)=0$ for every $j\in S$, that is, if, and only if, $i$ is
an ancestor of every labeled node.  Since the set of labeled nodes of  $T$
includes all leaves and all elementary nodes, this is equivalent to
the fact that $i$ is the root.

As far as (b) goes, assume that $W_i\neq\emptyset$, so that $i$ has a
parent $x$.  Let $w_i$ be the weight of the arc $(x,i)$.  Then, since
every non-trivial path $[i,j]_T\pathgr i$ must end with the arc $(x,i)$,
it is clear that if $\ell_T(i,j)>0$, then $\ell_T(i,j)\geq w_i$.

Now, if $x$ is labeled, say with label $i_0$, then $x=[i,i_0]_T$ and
thus $\ell_T(i,i_0)=w_i$.  If $x$ is not labeled, then it cannot be
elementary, and hence it must have at least another child $y$.  Let
$i_0$ be a descendant leaf of $y$.  In this case, $x=[i,i_0]_T$ and
$\ell_T(i,i_0)=w_i$, too.  This proves that, in all cases, $w_i\in
W_i$, and thus that it is the smallest element of this set.  \qed
\end{proof}

The following result is a direct consequence of the last two lemmas.

\begin{corollary}
\label{rem:prewtree}
Let $T$ and $T'$ be two $\RRp$-weighted phylogenetic trees on the same
set $S$ of taxa such that $\ell(T)=\ell(T')$.  Then:

\begin{enumerate}[(a)]
\item The nested taxa of $T$ and $T'$ are the same.

\item $T$ has its root labeled with $i$ if, and only if, $T'$ has its
root labeled with $i$.

\item If the nodes labeled with $i$ in $T$ and $T'$ are not their
roots, the weight of the arc with head $i$ in $T$ and in $T'$ is the
same.  \qed
\end{enumerate}
\end{corollary}

Let $S$ be a set of taxa and $\mathcal{R}(S)$ the set of
\emph{$S$-triples}, that is, of structures $ab|c$ with $a,b,c\in S$
pairwise different.  Classically, an $S$-triplet $ab|c$  is said to be \emph{present} in a phylogenetic tree $T$ if  $c$ diverged from $a$ before $b$ did, in the sense that $[a,b]_T<[a,c]_T=[b,c]_T$.
Let now $(T,\omega)$ be an $\RRz$-weighted
phylogenetic tree on $S$.  For every $ab|c\in \mathcal{R}(S)$, let
$\lambda_T(ab|c)\in \RRz$ be defined as follows:
\begin{itemize}
\item If $ab|c$ is present in $T$,  then $\lambda_T(ab|c)$ is
the distance from $[a,c]_T=[b,c]_T$ to $[a,b]_T$ 

\item  If $ab|c$ is not present in $T$,  then $\lambda_T(ab|c)=0$.
\end{itemize}
Notice that $\lambda_T(ab|c)=\lambda_T(ba|c)$.

This mapping $\lambda_T$ has a simple description in terms of $\ell(T)$.

\begin{lemma}
\label{lem:lambda}
Let $(T,\omega)$ be an $\RRz$-weighted phylogenetic tree on $S$.  For
every $ab|c\in \mathcal{R}(S)$,
$$
\lambda_{T}(ab|c)=\max\{\ell_T(a,c)-\ell_T(a,b),0\}.
$$
\end{lemma}

\begin{proof}
If $[a,c]_T$ is a non-trivial ancestor of $[a,b]_T$ in $T$, then the path
$[a,c]_T\pathgr a$ contains the node $[a,b]_T$ and the distance
$\ell_T(a,c)$ from $[a,c]_T$ to $a$ is equal to the distance
$\lambda_{T}(ab|c)$ from $[a,c]_T$ to $[a,b]_T$ plus the distance
$\ell_T(a,b)$ from $[a,b]_T$ to $a$.  Therefore, in this case,
$$
\max\{\ell_T(a,c)-\ell_T(a,b),0\}=\ell_T(a,c)-\ell_T(a,b)=\lambda_{T}(ab|c).
$$
If $[a,c]_T=[a,b]_T$, then $\ell_T(a,c)=\ell_T(a,b)$ and $ab|c$  is not present in $T$ and thus
$$
\max\{\ell_T(a,c)-\ell_T(a,b),0\}=0=\lambda_{T}(ab|c).
$$
Finally, if $[a,c]_T$ is not an ancestor of $[a,b]$, then it
must happen that $[a,b]_T$ is a non-trivial ancestor of $[a,c]_T$ and
therefore $\ell_T(a,b)\geq \ell_T(a,c)$. Since $ab|c$  is not present in $T$, either,  this implies that
$$
\max\{\ell_T(a,c)-\ell_T(a,b),0\}=0=\lambda_{T}(ab|c).
$$
So, the equality in the statement always holds.\qed
\end{proof}

The following result is Thm.~2 in \cite{moulton.ea:08}.  In it,
$\mathcal{Q}(X)$ denotes the set of $X$-\emph{quartets}, that is, of
structures $ab|cd$ with $a,b,c,d\in X$ pairwise different.

\begin{theorem}
\label{thm:moulton}
Let $\lambda:{\mathcal R}(S)\to \RRz$ be a map such that
$\lambda(ab|c)=\lambda(ba|c)$ for every $a,b,c\in S$ pairwise
different, and let $z$ be an element not in $S$.  Then:
\begin{enumerate}[(a)]
\item $\lambda=\lambda_T$ for some $\RRz$-weighted phylogenetic tree
$(T,\omega)$ with neither nested taxa nor weight 0 internal arcs if,
and only if, the mapping $\mu:\mathcal{Q}(S\cup\{z\}) \to \RRz$
defined by
$$
\mu(ab|cd)=\left\{
\begin{array}{ll}
\lambda(ab|c) & \mbox{ if $d=z$}\\
\mathrm{min}\{\lambda(ab|c),\lambda(ab|d)\}+\mathrm{min}\{\lambda(cd|a),\lambda(cd|b)\}
& \mbox{ if $d\neq z$}
\end{array}
\right.
$$
satisfies the following properties:
\begin{enumerate}[(1)]
\item $\mu(ab|cd)=\mu(ba|cd)=\mu(cd|ab)$

\item For every $a,b,c,d$, at least two of $\mu(ab|cd)$, $\mu(ac|bd)$,
and $\mu(ad|bc)$ are equal to 0.

\item If $\mu(ab|cd)>0$, then, for every $x\neq a,b,c,d$, either
$\mu(ab|cx)\cdot \mu(ab|dx)>0$ or $\mu(ax|cd)\cdot \mu(bx|cd)>0$.

\item For every $a,b,c,d,e$, if $\mu(ab|cd)>\mu(ab|ce)>0$, then
$$
\mu(ae|cd)=\mu(ab|cd)-\mu(ab|ce).
$$

\item For every $a,b,c,d,e$, if $\mu(ab|cd)>0$ and $\mu(bc|de)>0$, then
$$
\mu(ab|de)=\mu(ab|cd)+\mu(bc|de).
$$
\end{enumerate}

\item If $(T,\omega)$ and $(T',\omega')$ are two $\RRz$-weighted
phylogenetic trees with neither nested taxa nor weight 0 internal arcs
and such that $\lambda_T=\lambda_{T'}$, then $T\cong T'$ as
phylogenetic trees and the isomorphism preserves the weights of the
internal arcs.\qed
\end{enumerate}
\end{theorem}

Now we can proceed with the proof that splitted path lengths
matrices characterize $\RRp$-weighted phylogenetic trees.

\begin{theorem}
\label{prop:nod-arbtree}
Two $\RRp$-weighted phylogenetic trees on the same set $S$ of taxa are
isomorphic if, and only if, they have the same splitted path lengths
matrices.
\end{theorem}

\begin{proof}
As in Proposition \ref{prop:nod-bintree}, the statement when $|S|=1$ is obviously true.
Assume now that $|S|\geq 2$.  For every $\RRp$-weighted phylogenetic
tree $(T,\omega)$ on $S$, let $(\overline{T},\overline{\omega})$ be
the $\RRz$-weighted phylogenetic tree without nested taxa obtained as
follows: for every internal labeled node $i$ of $T$, unlabel it and
add to it a leaf child labeled with $i$ through an arc of weight 0.
It is straightforward to check that
$\ell_T(i,j)=\ell_{\overline{T}}(i,j)$ for every $i,j\in S$.  Since
$T$ was $\RRp$-weighted, the only weight 0 arcs in $\overline{T}$ are
the new pendant arcs that replace the nested taxa.  Moreover,
$(T,\omega)$ can be recovered from $(\overline{T},\overline{\omega})$
by simply removing the weight 0 pendant arcs and labeling the tail of
a removed arc with the label of the arc's head.

Let now $(T_1,\omega_1)$ and $(T_2,\omega_2)$ be two $\RRp$-weighted
phylogenetic trees on the same set $S$ of taxa such that
$\ell(T_1)=\ell(T_2)$.  Then
$\ell(\overline{T}_1)=\ell(\overline{T}_2)$ and hence, by Lemma
\ref{lem:lambda}, $\lambda_{\overline{T}_1}=\lambda_{\overline{T}_2}$.
Since $(\overline{T}_1,\overline{\omega}_1)$ and $(\overline{T}_2,
\overline{\omega}_2)$ are $\RRz$-weighted phylogenetic trees with
neither nested taxa nor weight 0 internal arcs, by Theorem
\ref{thm:moulton}.(b) we have that $\overline{T}_1\cong
\overline{T}_2$ as phylogenetic trees, and moreover this isomorphism
preserves the weights of the internal arcs.  But we also know that the
arc ending in the leaf $i$ has the same weight in $\overline{T}_1$ and
in $\overline{T}_2$: if $i$ was a nested taxon of $T_1$
and $T_2$ (and recall that $T_1$ and $T_2$ have the same nested taxa
by Corollary \ref{rem:prewtree}.(a)), this weight is in both cases 0,
and if $i$ was the label of a leaf of $T_1$ and $T_2$, this weight is
the same in $T_1$ and in $T_2$ by Corollary \ref{rem:prewtree}.(c),
and hence in $\overline{T}_1$ and in $\overline{T}_2$.

Therefore, the isomorphism $\overline{T}_1\cong \overline{T}_2$ is an
isomorphism of weighted phylogenetic trees.  Finally, the way
$(T_1,\omega_1)$ and $(T_2,\omega_2)$ are reconstructed from
$(\overline{T}_1,\overline{\omega}_1)$ and $(\overline{T}_2,
\overline{\omega}_2)$ implies that this isomorphism induces an
isomorphism of weighted phylogenetic trees $T_1\cong T_2$.

This proves the `if' implication; the `only if' implication is
obvious.  \qed
\end{proof}

\begin{remark}
The proof of the last theorem can also be applied, with small
modifications, to prove that the splitted path lengths matrices also
separate $\RRp$-weighted phylogenetic trees \emph{with multi-labeled nodes},
that is, where a node can have more than one label (but two different
nodes cannot share any label); in such a tree $T$,   if $i$ and $j$ are labels of the same node, then $\ell_T(i,j)=\ell_T(j,i)=0$.  It is enough to slightly change
the definition of $\overline{T}$: on the one hand, for every internal
labeled node of $T$, unlabel it and, for each one of its labels, add
to it a leaf child labeled with this label through an arc of weight 0;
{and}, on the other hand, do the same for every leaf with more than
one label. The same argument as in the proof of the last theorem shows that if $T_1$ and $T_2$ are two 
 $\RRp$-weighted phylogenetic trees with multi-labeled nodes such that $\ell(T_1)=\ell(T_2)$,  then
the $\RRz$-weighted
phylogenetic trees with neither nested taxa nor weight 0 internal arcs $\overline{T}_1$ and $\overline{T}_2$ obtained in this way are isomorphic. To derive from this isomorphism an isomorphism $T_1\cong T_2$,  one must use that, in this
multi-labeled case:
\begin{itemize}
\item An internal node of a tree $T$ is labeled $\{i_1,\ldots,i_k\}$ if, and only
if, $\ell_T(a,b)=0$ for every $a,b\in \{i_1,\ldots,i_k\}$,
$\ell_T(a,j)>0$ or $\ell_T(j,a)>0$ for every $a\in \{i_1,\ldots,i_k\}$ and every $j\notin
\{i_1,\ldots,i_k\}$, and there exists some 
$j\notin
\{i_1,\ldots,i_k\}$ such that $\ell_T(a,j)=0$  for every $a\in \{i_1,\ldots,i_k\}$.

\item A leaf of $T$ is labeled $\{i_1,\ldots,i_k\}$ if, and only
if, $\ell_T(a,b)=0$ for every $a,b\in \{i_1,\ldots,i_k\}$, and
$\ell_T(a,j)>0$ for every $a\in \{i_1,\ldots,i_k\}$ and every $j\notin
\{i_1,\ldots,i_k\}$.
\end{itemize}
These properties entail that if $\ell(T_1)=\ell(T_2)$, then $T_1$ and $T_2$ have the same families of sets $\{i_1,\ldots,i_k\}$ of labels of internal nodes as well as of leaves.
We leave the details to the reader. 
\end{remark}

Notice that Theorem \ref{thm:moulton} not only establishes that the
mapping $\lambda_T$ singles out an $\RRz$-weighted phylogenetic tree $T$
with neither nested taxa nor weight 0 internal arcs, up to the
weights of its pendant arcs, but it also characterizes what mappings
can be realized as $\lambda_T$-mappings, for some $T$ of this type.
We can use this result to characterize the matrices that are splitted
path lengths matrices of $\RRp$-weighted phylogenetic trees.

\begin{proposition}
Let $M=\big(m_{i,j}\big)\in \mathcal{M}_n(\RRz)$ be an $n\times n$
square matrix over $\RRz$ with $m_{i,i}=0$ for every $i=1,\ldots,n$.
Then, $M=\ell(T)$ for some $\RRp$-weighted phylogenetic tree $T$ on
$S=\{1,\ldots,n\}$ if, and only if, the mapping
$\lambda_M:\mathcal{R}(S)\to \RRz$ defined by
$$
\lambda_M(ab|c)=\max\{m_{a,c}-m_{a,b},0\}
$$
satisfies the following conditions:
\begin{enumerate}[(a)]
\item $\lambda_M(ab|c)=\lambda_M(ba|c)$ for every $a,b,c\in S$ pairwise different.

\item The mapping $\mu_M$ defined from $\lambda_M$ as in Theorem
\ref{thm:moulton}.(a) satisfies properties (1)--(5) therein.
\end{enumerate}
\end{proposition}

\begin{proof}
The `only if' implication is easy: if $M=\ell(T)$, so that
$m_{i,j}=\ell_T(i,j)$ for every $i,j\in S$, then
$\lambda_M=\lambda_{\overline{T}}$, with $\overline{T}$ the
$\RRz$-weighted phylogenetic tree without nested taxa or weight 0
internal arcs associated to $T$ in the proof of Theorem
\ref{prop:nod-arbtree}, and therefore it satisfies conditions (a) and
(b) in the statement.

Conversely, if $\lambda_M$ satisfies conditions (a) and (b), then by
Theorem \ref{thm:moulton} there exists an $\RRz$-weighted phylogenetic
tree $T_0$ without nested taxa or weight 0 internal arcs such that
$\lambda_M=\lambda_{T_0}$.  By Lemma \ref{lem:lambda},
$\lambda_{T_0}(ab|c)=\max\{\ell_{T_0}(a,c)-\ell_{T_0}(a,b),0\}$.
Therefore, for every $a,b,c\in S$ pairwise different,
$$
\max\{\ell_{T_0}(a,c)-\ell_{T_0}(a,b),0\}=\max\{m_{a,c}-m_{a,b},0\}.
$$
The tree $T_0$ is unique up to the weights of the pendant arcs.  So,
without any loss of generality we may assume that the weight of the
arc ending in the leaf $a$ is
$$
\min\{m_{a,j}\mid j\neq a\}.
$$

Now, for every $a\in S$ and for every $b\in S\setminus\{a\}$, $b$ is a
descendant of the parent $x_a$ of $a$ in $T_0$ if, and only if,
$m_{a,b}=\min\{m_{a,j}\mid j\neq a\}$.  As far as the `if' implication
goes, assume that $m_{a,b}=\min\{m_{a,j}\mid j\neq a\}$ but $b$ is not
a descendant of $x_a$.  Let $c\in S\setminus\{a\}$ be a descendant of $x_a$, so that
$[a,c]_{T_0}=x_a$.  Then, $[a,c]_{T_0}$ is a non-trivial descendant of $[a,b]_{T_0}$ and
therefore (since the internal arcs of $T_0$ have non-negative weight),
$\ell_{T_0}(a,b)-\ell_{T_0}(a,c)>0$.  But this contradicts the fact
that, since $m_{a,c}\geq m_{a,b}$,
$$
\ell_{T_0}(a,b)-\ell_{T_0}(a,c)=\lambda_{T_0}(ac|b)=\lambda_M(ac|b)=\min\{m_{a,b}-m_{a,c},0\}=0.
$$
As far as the converse implication goes, let $b\in S\setminus\{a\}$ be a descendant of
$x_a$, and let $b'\in S\setminus\{a\}$ be such that
$m_{a,b'}=\min\{m_{a,j}\mid j\neq a\}$: as we have just seen, $b'$ is
also a descendant of $x_a$ and therefore $[a,b]_{T_0}=[a,b']_{T_0}=x_a$.  Then,
$\max\{m_{a,b}-m_{a,b'},0\}=\lambda_{T_0}(ab'|b)=0$ implies that
$m_{a,b}-m_{a,b'}\leq 0$, that is, that $m_{a,b}=\min\{m_{a,j}\mid
j\neq a\}$, too.

Now, let us a fix a taxon $a\in S$, and let $b\in S\setminus\{a\}$ be
a descendant of the parent $x_a$ of $a$ in $T_0$.  Then, on the one
hand, $\ell_{T_0}(a,b)=m_{a,b}$, because it is the weight of the arc
$(x_a,a)$, and, on the other hand, for every $c\neq a,b$, we have that
$m_{a,c}\geq m_{a,b}$ 
and $\ell_{T_0}(a,c)\geq \ell_{T_0}(a,b)$
and therefore
$$
m_{a,c}=\lambda_M(ab|c)+m_{a,b}=\lambda_{T_0}(ab|c)+\ell_{T_0}(a,b)=\ell_{T_0}(a,c).
$$
This implies that the $a$-th row in $M$ and $\ell(T_0)$ are equal, and
hence, since $a$ was any element of $S$,  $M=\ell(T_0)$.

Finally, $T_0$ is transformed into an $\RRp$-weighted phylogenetic
tree with the same splitted path lengths matrix by simply removing the
weight 0 pendant arcs and labeling the tail of a removed arc with the
label of the arc's head; cf.  the proof of Theorem
\ref{prop:nod-arbtree}.  \qed
\end{proof}

\section{Splitted nodal metrics}

Let $\TT_n$ be the space of $\RRp$-weighted phylogenetic trees on the
set $S=\{1,\ldots,n\}$ of taxa.  As we have seen, the mapping
$$
\ell: \TT_n \longrightarrow \mathcal{M}_n(\RRz)
$$
that associates to each $(T,\omega)\in \TT_n$ its splitted path
lengths matrix $\ell(T)$ is injective up to isomorphisms.  As it
happened with the embedding $L:\bTT_n\hookrightarrow \RR^{n(n-1)/2}$,
this allows one to induce metrics on $\TT_n$ from metrics on
$\mathcal{M}_n(\RRz)$.

\begin{proposition}
Let $D$ be any metric on $\mathcal{M}_n(\RRz)$. The mapping
$$
\begin{array}{rcl}
d: \TT_n\times \TT_n & \to & \RRz\\
 (T_1,T_2) & \mapsto & D(\ell(T_1),\ell(T_2))
 \end{array}
 $$
 satisfies the axioms of metrics up to isomorphisms:
  \begin{enumerate}[(1)]
  \item $d(T_1,T_2)\ge 0$,
  \item $d(T_1,T_2)=0$ if, and only if, $T_1\cong T_2$,
  \item $d(T_1,T_2)=d(T_2,T_1)$,
  \item $d(T_1,T_3)\le d(T_1,T_2)+d(T_2,T_3)$.
  \end{enumerate}
 \end{proposition}

\begin{proof}
Properties (1), (3) and (4) are direct consequences of the
corresponding properties of $D$, while property (2) follows from the
separation axiom for $D$ (which says that $D(M_1,M_2)=0$ if, and only if, $M_1=M_2$)
and Theorem \ref{prop:nod-arbtree}.\qed
\end{proof}

We shall generically call \emph{splitted nodal metrics} the metrics on
$\TT_n$ induced by metrics on $\mathcal{M}_n(\RRz)$ through the
embedding $\ell$.  In particular, every $L^p$ norm $\|\, \cdot\, \|_p$
on $\mathcal{M}_n(\RRz)$ defines a splitted nodal metric $d_p^s$
through
$$
d_p^s(T_1,T_2)=\|\ell(T_1)-\ell(T_2)\|_p.
$$
For instance,
$$
\begin{array}{l}
d^s_1(T_1,T_2)=\displaystyle \sum_{1\leq i\neq j\leq
n}|\ell_{T_1}(i,j)-\ell_{T_2}(i,j)|,\\
d^s_2(T_1,T_2)= \displaystyle\sqrt{\sum_{1\leq i\neq j\leq
n}(\ell_{T_1}(i,j)-\ell_{T_2}(i,j))^2}
\end{array}
$$
are the splitted nodal metrics induced by the $L^1$ and $L^2$ norms on
$\mathcal{M}_n(\RRz)$.

We have seen in the previous section that the splitted path lengths matrices can be computed in $O(n^2)$ time. Their difference can be computed 
in $O(n^2)$ time, and the sum of the $p$-th powers of the entries of the resulting matrix can be computed in $O(n^2\log(p)+n^2)$ time  (assuming constant-time addition and multiplication of real numbers). Therefore, the cost of computing $d_p^s(T_1,T_2)^p$, for $T_1,T_2\in \TT_n$ and $p\in \NN^+$, is $O(n^2\log(p)+n^2)$. Thus, if $p=1$, the $d_1^s$ metric on $ \TT_n$ can be computed in $O(n^2)$ time. For $p\geq 2$, the cost of computing $d_p^s(T_1,T_2)$,  for $T_1,T_2\in \TT_n$, as the $p$-th root of $d_p^s(T_1,T_2)^p$ will depend on the accuracy with which this root is computed. For instance,
using the Newton method to compute it with an accuracy of an $1/2^h$-th of its value
has a cost of $O(p^2\log(p)\log(hp))$; see, for instance, \cite{batra:2008}.  So, in practice, for small $p$ and not too large $h$,  this step will be dominated by the computation of 
$d_p^s(T_1,T_2)^p$, and  the total cost will be  $O(n^2)$ (we understand in this case $\log(p)$ as part of the constant factor).
For $p=0$ or $\infty$, the cost of computing 
$d_p(T_1,T_2)$  is also $O(n^2)$ time.

These splitted nodal metrics can be seen conceptually as the
generalizations to $\TT_n$ of the classical nodal metrics on $\bTT_n$.
Conceptually, but not numerically, because the restriction of $d_p^s$
to $\bTT_n$ is not equal to $d_p$, even up to a scalar factor, as the
following easy example shows.

\begin{example}
\label{ex:dpnodps}
Consider the non-weighted binary trees $T_1,T_2,T_3$ depicted in
Fig.~\ref{fig:dpnodps}.
\begin{figure}[htb]
\begin{center}
            \begin{tikzpicture}[thick,>=stealth,scale=0.3]
              \draw(0,0) node[tre] (1) {}; \etq 1
              \draw(2,0) node[tre] (2) {}; \etq 2
              \draw(4,0) node[tre] (3) {}; \etq 3
              \draw(6,0) node[tre] (4) {}; \etq 4
              \draw(5,2) node[tre] (b) {};  
              \draw(3,4) node[tre] (a) {};  
              \draw(1,6) node[tre] (r) {};  
             \draw[->] (r)--(1);
            \draw[->] (r)--(a);
             \draw[->] (a)--(b);
                     \draw[->] (a)--(2);
               \draw[->] (b)--(3);
             \draw[->] (b)--(4);
    \draw(3,-1.5) node  {$T_1$};
             \end{tikzpicture}
  \qquad
  \begin{tikzpicture}[thick,>=stealth,scale=0.3]
               \draw(0,0) node[tre] (1) {}; \etq 1
              \draw(2,0) node[tre] (2) {}; \etq 2
              \draw(4,0) node[tre] (3) {}; \etq 3
              \draw(6,0) node[tre] (4) {}; \etq 4
              \draw(1,2) node[tre] (b) {};  
              \draw(3,4) node[tre] (a) {};  
              \draw(5,6) node[tre] (r) {};  
             \draw[->] (r)--(4);
            \draw[->] (r)--(a);
             \draw[->] (a)--(b);
                     \draw[->] (a)--(3);
               \draw[->] (b)--(1);
             \draw[->] (b)--(2);
    \draw(3,-1.5) node  {$T_2$};
   \end{tikzpicture}
   \qquad
           \begin{tikzpicture}[thick,>=stealth,scale=0.3]
              \draw(0,0) node[tre] (1) {}; \etq 1
              \draw(2,0) node[tre] (4) {}; \etq 4
              \draw(4,0) node[tre] (3) {}; \etq 3
              \draw(6,0) node[tre] (2) {}; \etq 2
              \draw(5,2) node[tre] (b) {};  
              \draw(3,4) node[tre] (a) {};  
              \draw(1,6) node[tre] (r) {};  
             \draw[->] (r)--(1);
            \draw[->] (r)--(a);
             \draw[->] (a)--(b);
                     \draw[->] (a)--(4);
               \draw[->] (b)--(3);
             \draw[->] (b)--(2);
    \draw(3,-1.5) node  {$T_3$};
             \end{tikzpicture}
    \end{center}
\caption{\label{fig:dpnodps} 
The non-weighted binary phylogenetic trees in Example \ref{ex:dpnodps}.}
\end{figure}
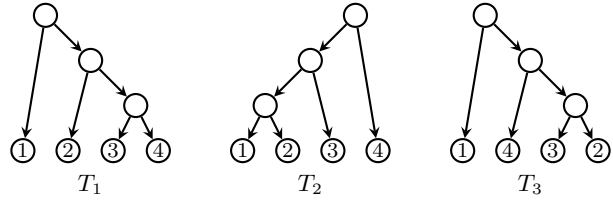
It is easy to compute their path lengths vectors and splitted path
lengths matrices:
$$
\begin{array}{c}
L(T_1)=(3,4,4,3,3,2),\
L(T_2)=(2,3,4,3,4,3),\
L(T_3)=(4,4,3,2,3,3)\\[2ex]
\ell(T_1)=\left(\begin{array}{llll}
0 & 1 & 1 & 1 \\[-0.5ex]
2 & 0 & 1 & 1 \\[-0.5ex]
3 & 2 & 0 & 1 \\[-0.5ex]
3 & 2 & 1 & 0
\end{array}\right),
\
\ell(T_2)=\left(\begin{array}{llll}
0 & 1 & 2 & 3 \\[-0.5ex]
1 & 0 & 2 & 3 \\[-0.5ex]
1 & 1 & 0 & 2 \\[-0.5ex]
1 & 1 & 1 & 0
\end{array}\right),
\
\ell(T_3)=\left(\begin{array}{llll}
0 & 1 & 1 & 1 \\[-0.5ex]
3 & 0 & 1 & 2 \\[-0.5ex]
3 & 1 & 0 & 2 \\[-0.5ex]
2 & 1 & 1 & 0
\end{array}\right).
\end{array}
$$
From these vectors and matrices we obtain that
$$
d_p(T_1,T_2)=d_p(T_1,T_3)=\left\{
\begin{array}{ll}
4 & \mbox{ if $p=0$}\\
\sqrt[p]{4} & \mbox{ if $p\in \NN^+$}\\
1& \mbox{ if $p=\infty$}
\end{array}
\right.
$$
while
$$
d_p^s(T_1,T_2)=\left\{
\begin{array}{ll}
10 & \mbox{ if $p=0$}\\
\sqrt[p]{6+4\cdot 2^p} & \mbox{ if $p\in \NN^+$}\\
2& \mbox{ if $p=\infty$}
\end{array}
\right.
\qquad
d_p^s(T_1,T_3)=\left\{
\begin{array}{ll}
6 & \mbox{ if $p=0$}\\
\sqrt[p]{6} & \mbox{ if $p\in \NN^+$}\\
1 & \mbox{ if $p=\infty$}
\end{array}
\right.
$$
This shows that there does not exist any $\lambda\in \RR$ such that
$d_p^s=\lambda\cdot d_p$ on $\bTT_4$ for any $p\in \NN\cup\{\infty\}$.
Similar counterexamples can be produced for every $n\geq 4$.
\end{example}

The following inequality relates $d_p$ and $d_p^s$ on any
$\bTT_n$.

 \begin{proposition}
For every $T_1,T_2\in \bTT_n$ and for every $p\in \NN\cup \{\infty\}$,
$$
d_p(T_1,T_2)\leq 
\left\{
\begin{array}{ll}
d_p^s(T_1,T_2)  & \mbox{if $p=0$}\\
2^{1-\frac{1}{p}} d_p^s(T_1,T_2)& \mbox{if $p\in \NN^+$}\\
2 d_p^s(T_1,T_2)& \mbox{if $p=\infty$}
\end{array}
\right.
$$
\end{proposition}

\begin{proof}
For every $T\in \bTT_n$, let $L^*(T)$ be the symmetric matrix
$$
L^*(T)=\ell(T)+\ell(T)^t.
$$
Notice that the $(i,j)$-th and the $(j,i)$-th entries of $L^*(T)$ are
both equal to $L_T(i,j)$.  Now, by the usual properties of norms,
$$
\begin{array}{rl}
\|L^*(T_1)-L^*(T_2)\|_p & =
\|\ell(T_1)+\ell(T_1)^t-(\ell(T_2)+\ell(T_2)^t)\|_p\\ & \leq
\|\ell(T_1)-\ell(T_2)\|_p+\|\ell(T_1)^t-\ell(T_2)^t\|_p\\ & =
2 \|\ell(T_1)-\ell(T_2)\|_p.
\end{array}
$$
On the other hand, $L^*(T_1)-L^*(T_2)$ can be understood as two
concatenated copies of $L(T_1)-L(T_2)$ and therefore,
$$
\|L^*(T_1)-L^*(T_2)\|_p=\left\{
\begin{array}{ll}
2\|L(T_1)-L(T_2)\|_p  & \mbox{if $p=0$}\\
\sqrt[p]{2}\cdot \|L(T_1)-L(T_2)\|_p& \mbox{if $p\in \NN^+$}\\
\|L(T_1)-L(T_2)\|_p& \mbox{if $p=\infty$}
\end{array}
\right.
$$
Combining this equality with the previous inequality we obtain the
inequality in the statement.  \qed
\end{proof}

\section{The non-weighted case}

Although weights enrich the topological structure of a phylogenetic
tree, for instance by adding probabilities, bootstrap values or divergence degrees to branches, the comparison of non-weighted phylogenetic trees, as
bare hierarchical classifications  or evolutive histories, has an interest in itself.  
Let  $\nTT_n$ denote the class of all non-weighted
phylogenetic trees on $S=\{1,\ldots,n\}$.
Felsenstein \cite{fel:78} gave a recurrent formula for the number $U(n,m)$ of different trees in $\nTT_n$ with $m$ unlabeled internal nodes, from which the total number $|\nTT_n|$ of different non-weighted phylogenetic trees on $n$ taxa can be computed: see Table 2 in  \cite{fel:78} or sequence A005264 in \cite{OEIS}. Table \ref{table:fel} recalls the first values of $|\nTT_n|$.

\begin{table}[htdp]
\centering
\begin{tabular}{r||c|c|c|c|c|c|c|c|}
$n$ & 1 & 2 & 3 & 4 & 5 & 6 & 7 \\
\hline
$|\nTT_n|$ & 1 & 3 & 22 & 262 & 4\,336 & 91\,984 & 2\,381\,408
\end{tabular}

\caption{\label{table:fel}The values of $|\nTT_n|$ for $n$ up to 7}
\end{table}%

In
this section we gather some results on the splitted
nodal metrics $d_p^s$, for $p\in\NN^+$, on $\nTT_n$, and we report on some numerical experiments
for $d_1^s$ and $d_2^s$ on this class. 
To simplify the notations, for every $a,b\in S$ and $p\in \NN^+$, we
shall write $C_{T_1,T_2}^p(a,b)$ to denote
$|\ell_{T_1}(a,b)-\ell_{T_2}(a,b)|^p$.  In this way, if $T_1,T_2\in
\nTT_n$ and $p\in \NN^+$, then
$$
d_p^s(T_1,T_2)^p=\sum_{(a,b)\in S^2} C_{T_1,T_2}^p(a,b)\in\NN.
$$

Our first result shows that the metrics $d_p^s$ have a redundant factor on $\nTT_n$ when $n$ is odd.
 
 \begin{lemma}
 \label{lem:evennorm}
If $n$ is odd, then $\|\ell(T)\|_1$ is even, for every $T\in \nTT_n$.
\end{lemma}

\begin{proof}
Let $T=(E,V)$ be a non-weighted phylogenetic tree on
$S=\{1,\ldots,n\}$ with $n$ odd.  For every $e\in E$, let
$\nu_\ell(e)$ be the number of paths $[i,j]\pathgr i$, with $i,j\in S$,  that contain the arc
$e$.  It is clear that
$$
\|\ell(T)\|_1=\sum_{1\leq i\neq j\leq n} \ell_T(i,j)=\sum_{e\in E} \nu_\ell(e).
$$
It turns out that if $n$ is odd, then every $\nu_\ell(e)$ is even and
therefore the right-hand side sum is even.  Indeed, let $e=(u,v)$ be
any arc and let $V$ be the set of descendant labeled nodes of $v$.
Then, $e$ is contained in a path $[i,j]\pathgr i$ if, and only if,
$i\in V$ and $j\notin V$.  This shows that $\nu_\ell(e)=|V|\cdot
|S-V|$.  Now, since $|S|$ is odd, either $|V|$ or $|S-V|$ is even,
which implies that $\nu_\ell(e)$ is even.\qed
\end{proof}

\begin{proposition}
If $n$ is odd, then $d_p^s(T_1,T_2)^p$ is even, for every $T_1,T_2\in
\nTT_n$ and for every $p\in \NN^+$.
\end{proposition}

\begin{proof}
Let $T_1,T_2\in \nTT_n$, with $n$ odd. Then
$$
d_p^s(T_1,T_2)^p=\sum_{1\leq i\neq j\leq n} C_{T_1,T_2}^p(i,j).
$$
Now, we know that $\sum_{1\leq i\neq j\leq n} \ell_{T_1}(i,j)$ and
$\sum_{1\leq i\neq j\leq n} \ell_{T_2}(i,j)$ are even numbers.  This
implies that the number
$$
\big|\{(i,j)\in S^2\mid C_{T_1,T_2}^p(i,j)\mbox{ odd}\}\big|
=\big|\{(i,j)\in S^2\mid \ell_{T_1}(i,j)-\ell_{T_2}(i,j)\mbox{ odd}\}\big|
$$
is even, and hence that the sum $\sum_{1\leq i\neq j\leq
n}C_{T_1,T_2}^p(i,j)$ is even.\qed
\end{proof}

This result shows that if $n$ is odd, $d_1^s$ takes only even values on $\nTT_n$, and
therefore it can be divided by 2 and the resulting values are
still integer numbers.  In a similar way, $d_2^s$ has a `redundant'
$\sqrt{2}$ factor on $\nTT_n$, for $n$ odd. No similar result holds for even values of $n$: for instance, $\nTT_2$ consists of three trees $T_1,T_2,T_3$, with Newick strings \texttt{(1,2);}, \texttt{((1)2);}, and \texttt{((2)1);}, respectively, and
$d_1^s(T_1,T_2)=d_1^s(T_1,T_3)=1$, $d_1^s(T_2,T_3)=2$.

\begin{remark}
The theses in the last two results are true in the more general
setting of $\NN^+$-weighted phylogenetic trees.  To see it, notice that
if $(T,\omega)$ is such a tree, then
$$
\|\ell(T)\|_1=\sum_{1\leq i\neq j\leq n} \ell_T(i,j)=\sum_{e\in E} \omega(e)\cdot \nu_\ell(e)
$$
and then, the proof that each $\nu_\ell(e)$ is even is the same as in
the non-weighted case.

On the other hand, the thesis in the last proposition does not generalize to $p=0$ or $\infty$: it is easy to produce counterexamples showing
that $d_0^s$ and $d_\infty^s$ take odd values on $\nTT_3$.
\end{remark}

Our next goal is to find the least value for $d_p^s$ on $\nTT_n$, for
$p\in \NN^+$.

\begin{lemma}
\label{lem:5}
Let $T_1,T_2\in \nTT_n$ with $n\geq 6$ and $p\in \NN^+$. If there is some taxon that is a leaf of largest depth in $T_1$ but not in $T_2$, then $d_p(T_1,T_2)^p\geq 5$.
\end{lemma}

\begin{proof}
To simplify the notations, and since in this proof the trees
$T_1,T_2$ and the index $p$ are fixed, we shall write $C(a,b)$ to
denote $C_{T_1,T_2}^p(a,b)$. 

Assume, without any loss of generality, that $1$ is a deepest leaf of $T_1$ and that $2$ is a leaf of $T_2$ such that $\mathrm{depth}_{T_2}(2)>\mathrm{depth}_{T_2}(1)$. Then, the distance from $[1,2]_{T_2}$ to $2$ will be larger than to $1$. This implies that
$\ell_{T_2}(2,1)>\ell_{T_2}(1,2)$. Since $\ell_{T_1}(2,1)\leq \ell_{T_1}(1,2)$ (because $\mathrm{depth}_{T_1}(2)\leq \mathrm{depth}_{T_1}(1)$), it must happen that
$\ell_{T_2}(2,1)\neq \ell_{T_1}(2,1)$ or $\ell_{T_2}(1,2)\neq \ell_{T_1}(1,2)$, and therefore
$$
C(1,2)+C(2,1)\geq 1.
$$
Let us check now that, for every $a\in S\setminus\{1,2\}$, at least one of the following four equalities does not hold:
\begin{equation}\label{eq:1}
\begin{array}{l}
\ell_{T_2}(1,a)=\ell_{T_1}(1,a),\quad \ell_{T_2}(2,a)=\ell_{T_1}(2,a)
\\
\ell_{T_2}(a,1)=\ell_{T_1}(a,1), \quad \ell_{T_2}(a,2)=\ell_{T_1}(a,2)
\end{array}
\end{equation}
This will imply that every $a\in S\setminus\{1,2\}$
\emph{contributes} 1 to $d_p^s(T_1,T_2)^p$, in the sense that
$$
C(1,a)+C(2,a)+C(a,1)+C(a,2)\geq 1.
$$
Since there are at least 4 taxa in $S\setminus\{1,2\}$ and these contributions add up to 
$C(1,2)+C(2,1)$, this will prove that $d_p^s(T_1,T_2)^p\geq 5$.

The way each $a\in S\setminus\{1,2\}$ contributes to $d_p^s(T_1,T_2)^p$ depends on its relative position with respect to $1$ and $2$ in $T_2$.
\begin{itemize}
\item If $a\leq 1$, then $\ell_{T_2}(1,a)=0$ but $\ell_{T_1}(1,a)>0$ and therefore
$\ell_{T_2}(1,a)\neq \ell_{T_1}(1,a)$.

\item Assume that  $[a,1]_{T_2}=[a,2]_{T_2}\geq [1,2]_{T_2}$. In this case
$\ell_{T_2}(a,2)=\ell_{T_2}(a,1)$ and $\ell_{T_2}(2,a)>\ell_{T_2}(1,a)$. But these relations cannot hold in $T_1$, because they imply that $\mathrm{depth}_{T_1}(2)> \mathrm{depth}_{T_1}(1)$.  Thus,  the equalities (\ref{eq:1}) cannot hold simultaneously.

\item Assume that $1<[a,1]_{T_2}<[1,2]_{T_2}$. In this case $\lambda_{T_2}(a1|2)>0$ and
$$
\begin{array}{l}
\ell_{T_2}(a,1)+\lambda_{T_2}(a1|2)=\ell_{T_2}(a,2)\\
\ell_{T_2}(1,a)+\lambda_{T_2}(a1|2)=\ell_{T_2}(1,2)\\
\ell_{T_2}(2,a)=\ell_{T_2}(2,1)
\end{array}
$$
If $\ell_{T_1}(a,1)=\ell_{T_2}(a,1)$ and $\ell_{T_1}(a,2)=\ell_{T_2}(a,2)$, then the fact that
$\ell_{T_1}(a,2)>\ell_{T_1}(a,1)$
implies that
$1<[a,1]_{T_1}<[1,2]_{T_1}$  and thus
$$
\begin{array}{rl}
\lambda_{T_1}(a1|2) & =\ell_{T_1}(a,2)-\ell_{T_1}(a,1)\\ & =\ell_{T_2}(a,2)-\ell_{T_2}(a,1)=
\lambda_{T_2}(a1|2).
\end{array}
$$
Then, if $\ell_{T_1}(1,a)=\ell_{T_2}(1,a)$, 
$$
\begin{array}{rl}
\ell_{T_1}(1,2) & =\ell_{T_1}(1,a)+\lambda_{T_1}(a1|2)\\ & =\ell_{T_2}(1,a)+\lambda_{T_2}(a1|2)=\ell_{T_2}(1,2).
\end{array}
$$
Finally, if $\ell_{T_1}(2,a)=\ell_{T_2}(2,a)$, then
$$
\ell_{T_1}(2,1)=\ell_{T_1}(2,a)=\ell_{T_2}(2,a)=\ell_{T_2}(2,1).
$$
And this leads to a contradiction, because, as we have seen at the beginning of the proof,
$\ell_{T_2}(2,1)\neq \ell_{T_1}(2,1)$ or $\ell_{T_2}(1,2)\neq \ell_{T_1}(1,2)$. Therefore, the equalities (\ref{eq:1}) cannot hold simultaneously.

\item If $2<[a,2]_{T_2}<[1,2]_{T_2}$, a similar argument shows that at least one of  the equalities (\ref{eq:1}) fails, too.
%
%{\footnotesize (For authors' proofreading purposes only:
%In this case,$\lambda_{T_2}(a1|2)>0$ and
%$$
%\begin{array}{l}
%\ell_{T_2}(a,2)+\lambda_{T_2}(a2|1)=\ell_{T_2}(a,1)\\
%\ell_{T_2}(2,a)+\lambda_{T_2}(a2|1)=\ell_{T_2}(2,1)\\
%\ell_{T_2}(1,a)=\ell_{T_2}(1,2)
%\end{array}
%$$
%If $\ell_{T_1}(a,1)=\ell_{T_2}(a,1)$ and $\ell_{T_1}(a,2)=\ell_{T_2}(a,2)$, then
%$2<[a,2]_{T_1}<[1,2]_{T_1}$  and thus
%$$
%\begin{array}{rl}
%\lambda_{T_1}(a2|1) & =\ell_{T_1}(a,1)-\ell_{T_1}(a,2)\\ & =\ell_{T_2}(a,1)-\ell_{T_2}(a,2)=
%\lambda_{T_2}(a2|1).
%\end{array}
%$$
%If, moreover, $\ell_{T_1}(2,a)=\ell_{T_2}(2,a)$, 
%$$
%\begin{array}{rl}
%\ell_{T_1}(2,1) & =\ell_{T_1}(2,a)+\lambda_{T_1}(a2|1)\\ & =\ell_{T_2}(2,a)+\lambda_{T_2}(a2|1)=\ell_{T_2}(2,1).
%\end{array}
%$$
%Finally, if $\ell_{T_1}(1,a)=\ell_{T_2}(1,a)$, then
%$$
%\ell_{T_1}(1,2)=\ell_{T_1}(1,a)=\ell_{T_2}(1,a)=\ell_{T_2}(1,2).
%$$
%And this leads to the same contradiction as before, implying that the equalities (\ref{eq:1}) cannot hold simultaneously in this case, either.)
%}
\end{itemize}
This finishes the proof of the lemma.\qed
\end{proof}

\begin{figure}[htb]
\begin{center}
            \begin{tikzpicture}[thick,>=stealth,scale=0.3]
              \draw(0,0) node[tre] (1) {}; \etq 1
              \draw(2,2) node[tre] (2) {}; \etq 2
              \draw(4,4) node[tre] (r) {}; %\etq r
              \draw(2,0) node[tre] (3) {}; \etq 3
              \draw(4,0) node[tre] (4) {}; \etq 4
            \draw(6,0) node  {$\ldots$};
            \draw(8,0) node[tre] (n) {}; \etq n
                 \draw[->] (r)--(2);
                 \draw[->] (r)--(3);
                 \draw[->] (r)--(4);
                 \draw[->] (r)--(n);
                 \draw[->] (2)--(1);
           \draw(4,-1.5) node  {$T$};
            \end{tikzpicture}
  \qquad\qquad
%          \begin{tikzpicture}[thick,>=stealth,scale=0.3]
%              \draw(0,0) node[tre] (1) {}; \etq 1
%              \draw(2,2) node[tre] (3) {}; \etq 3
%              \draw(4,4) node[tre] (r) {}; %\etq r
%              \draw(2,0) node[tre] (2) {}; \etq 2
%              \draw(4,0) node[tre] (4) {}; \etq 4
%            \draw(6,0) node  {$\ldots$};
%            \draw(8,0) node[tre] (n) {}; \etq n
%                 \draw[->] (r)--(2);
%                 \draw[->] (r)--(3);
%                 \draw[->] (r)--(4);
%                 \draw[->] (r)--(n);
%                 \draw[->] (3)--(1);
%           \draw(4,-1.5) node  {$T_2$};
%            \end{tikzpicture}
            \begin{tikzpicture}[thick,>=stealth,scale=0.3]
              \draw(0,0) node[tre] (1) {}; \etq 1
              \draw(2,2) node[tre] (2) {}; \etq 2
              \draw(4,4) node[tre] (r) {}; %\etq r
              \draw(2,0) node[tre] (3) {}; \etq 3
              \draw(4,0) node[tre] (4) {}; \etq 4
            \draw(6,0) node  {$\ldots$};
            \draw(8,0) node[tre] (n) {}; \etq n
                 \draw[->] (r)--(2);
                 \draw[->] (r)--(3);
                 \draw[->] (r)--(4);
                 \draw[->] (r)--(n);
                 \draw[->] (3)--(1);
           \draw(4,-1.5) node  {$T'$};
            \end{tikzpicture}
\end{center}
\caption{\label{fig:dist4} 
Two non-isomorphic phylogenetic trees in $\nTT_n$ such that
$d_p^s(T,T')^p=4$ for every $p\in \NN^+$.}
\end{figure}
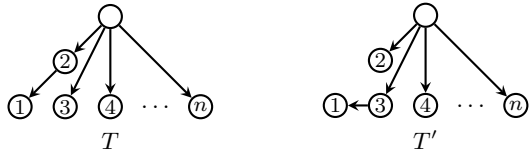

\begin{theorem}
For every $p\in \NN^+$ and for every $n\geq 2$:
\begin{enumerate}[(1)]
\item
If $n\leq 5$, then $\min\{d_p^s(T_1,T_2)^p\mid T_1,T_2\in \nTT_n,\
T_1\neq T_2\}=n-1$.

\item
If $n\geq 6$, then
$\min\{d_p^s(T_1,T_2)^p\mid T_1,T_2\in \nTT_n,\ T_1\neq T_2\}=4$.
\end{enumerate}
\end{theorem}

\begin{proof}
To simplify the notations, and since in this proof the trees
$T_1,T_2$ and the index $p$ are fixed, we shall write $C(a,b)$ to
denote $C_{T_1,T_2}^p(a,b)$. 

The cases $n=1$ to 5 can be checked `by hand' through the computation
of the distances between all pairs of trees in $\nTT_n$. In the case $n=1$, there is only one tree in $\nTT_1$, and, as we mentioned after Lemma~\ref{lem:evennorm},
$\nTT_2$ consists only of three trees $T_1,T_2,T_3$, with Newick strings \texttt{(1,2);}, \texttt{((1)2);}, and \texttt{((2)1);}, respectively, and it can be seen that
$d_p^s(T_1,T_2)^p=d_p^s(T_1,T_3)^p=1$, $d_p^s(T_2,T_3)^p=2$.
As far as the cases $n=3,4,5$ go, the files \texttt{\{3,4,5\}-tree-nt-pairs.dat} available at the Supplementary
Material web page contain the values of $d_p^s(T_1,T_2)^p$ for each (unordered) pair of trees $\{T_1,T_2\}$ in the corresponding $\nTT_n$.

Now, for $n\geq 5$, we shall prove by induction on $n$ that
$d_p^s(T_1,T_2)^p\geq 4$ for every pair of different trees $T_1,T_2\in
\nTT_n$.  Since it is easy to produce pairs of trees $T_1,T_2\in
\nTT_n$ such that $d_p^s(T_1,T_2)^p=4$, like for instance those
depicted in Fig.~\ref{fig:dist4}, this will finish the proof of the
statement.

The starting point for the induction procedure is $n=5$: we know (by
direct inspection of the file  \texttt{5-tree-nt-pairs.dat}) that $d_p^s(T_1,T_2)^p\geq 4$ for every pair of
different trees $T_1,T_2\in \nTT_5$.  Assume now that this inequality
holds for every two trees in $\nTT_n$, for some $n\geq 5$, and let us
prove it for $\nTT_{n+1}$.
 
So, let $T_1,T_2\in \nTT_{n+1}$ be a pair of different trees.  
As in the last proof, we shall write $C(a,b)$ to
denote $C_{T_1,T_2}^p(a,b)$.

Without any loss of generality, we assume that $n+1$ is a leaf of largest depth in
$T_1$.  By Lemma~\ref{lem:5}, if $n+1$ is not a deepest leaf of $T_2$, then $d_p^s(T_1,T_2)^p\geq 5$.
So, in the rest of the proof we assume that $n+1$ is also a deepest leaf of  $T_2$. 
In particular,  in both trees, the siblings of $n+1$ (if
they exist) are also deepest leaves.

We distinguish now two main cases, each
one divided in several subcases.

\begin{enumerate} 
\item[\textbf{\emph{(a)}}] Assume that  the parent of $n+1$ in $T_1$ is labeled, say with $n$.
This implies that
$$
\begin{array}{l}
\ell_{T_1}(n,n+1)=0,\ \ell_{T_1}(n+1,n)=1\\
\ell_{T_1}(n+1,a)=\ell_{T_1}(n,a)+1,\quad\mbox{for every $a\in S\setminus\{n,n+1\}$}\\ 
\ell_{T_1}(a,n+1)=\ell_{T_1}(a,n),\quad\mbox{for every $a\in S\setminus\{n,n+1\}$}
\end{array}
$$
We distinguish the following subcases.

\item[\emph{(a.1)}] Assume that, in $T_2$, the node $n$ is an ancestor of
$n+1$, but not its parent.  In this case,  $\ell_{T_2}(n+1,n)>1$,
and therefore
$$
C(n+1,n)\geq 1.
$$
Now, let $a\in S\setminus\{n,n+1\}$.  Let us see that $a$
{contributes} at least 1 to $d_p^s(T_1,T_2)^p$.
\begin{itemize}
\item If $n>[a,n+1]_{T_2}$  (that is, if $a$ is a descendant of an intermediate node in the path $n\pathgr n+1$), then 
$\ell_{T_2}(a,n+1)<\ell_{T_2}(a,n)$
and therefore, since $\ell_{T_1}(a,n+1)=\ell_{T_1}(a,n)$, it must
happen that $\ell_{T_1}(a,n+1)\neq \ell_{T_2}(a,n+1)$ or
$\ell_{T_1}(a,n)\neq \ell_{T_2}(a,n)$, which implies that
$$
C(a,n)+C(a,n+1)\geq 1.
$$

\item If $n\leq [a,n+1]_{T_2}$ in $T_2$, then
$$
\ell_{T_2}(n+1,a)=\ell_{T_2}(n,a)+\ell_{T_2}(n+1,n)>\ell_{T_2}(n,a)+1,
$$
and therefore, since $\ell_{T_1}(n+1,a)=\ell_{T_1}(n,a)+1$, 
 it must happen that
$\ell_{T_1}(n+1,a)\neq \ell_{T_2}(n+1,a)$ or $\ell_{T_1}(n,a)\neq \ell_{T_2}(n,a)$, and hence
$$
C(n,a)+C(n+1,a)\geq 1.
$$
\end{itemize}
Since  there are at least 4 taxa other than $n$ and
$n+1$, and their contributions add up to $C(n+1,n)$, we conclude that, in this case, $d_p^s(T_1,T_2)^p\geq 5$.

\item[\emph{(a.2)}] Assume that, in $T_2$, the node $n$ is not an ancestor of $n+1$; set
$$
\ell_{T_2}(n,n+1)=x\geq 1,\  \ell_{T_2}(n+1,n)=y\geq 1.
$$
If $x\geq y$, then $\mathrm{depth}_{T_2}(n)\geq \mathrm{depth}_{T_2}(n+1)$ and thus, since $n+1$  was a deepest leaf of $T_2$,   $n$ would  also be a deepest leaf of $T_2$. But $n$ is not a deepest leaf of $T_1$ and therefore, in this case, we already know by Lemma \ref{lem:5} that $d_p^s(T_1,T_2)^p\geq 5$.

Assume now that $x<y$. Then, $y\geq 2$ and thus, on the one hand,
$$
C(n+1,n)+C(n,n+1)=(y-1)^p+x^p\geq 2
$$
and, on the other hand, the path $[n+1,n]_{T_2}\pathgr n+1$   has at least one
intermediate node: let $a_0\neq n+1$ be a labeled node that is a descendant of
the parent of $n+1$ (notice that, in this case, $a_0$ is either the parent of $n+1$ or its sibling).  Then,
$$
\ell_{T_2}(a_0,n+1)<\ell_{T_2}(a_0,n),\quad
\ell_{T_2}(n+1,a_0)=1\leq \ell_{T_2}(n,a_0)
$$
imply that 
$$
C(a_0,n+1)+C(a_0,n)\geq 1,\quad
C(n+1,a_0)+C(n,a_0)\geq 1.
$$
So, in this case, $d_p^s(T_1,T_2)\geq 4$.

\item[\emph{(a.3)}] Assume that, in $T_2$, the node $n+1$ is a leaf and its
parent is $n$. 
 Let $T_1^*,T_2^*\in \nTT_n$ be the trees obtained from $T_1$ and $T_2$, respectively,   by removing the leaf $n+1$ together with its pendant arc.  After
this operation, we have that, for every $1 \leq a\neq b\leq n$,
$\ell_{T_i^*}(a,b)= \ell_{T_i}(a,b)$ and therefore,
$C(a,b)=C_{T_1^*,T_2^*}^p(a,b)$. 
Then,
$$
d_p^s(T_1,T_2)^p \geq \displaystyle\sum _{1\leq a\neq b\leq n} C(a,b)=
 \sum _{1\leq a\neq b\leq n} C_{T_1^*,T_2^*}^p(a,b)=d_p^s(T_1^*,T_2^*)^p\geq 4,
$$
the last inequality being given by the induction hypothesis.

\item[\textbf{\emph{(b)}}] Assume now that the parent of $n+1$ is not labeled.  Therefore,
$n+1$ must have at least one sibling, which, we recall, is a leaf.
Without any loss of generality we assume that $n$ is a sibling of
$n+1$.  In this case, we have that
$$
\begin{array}{l}
\ell_{T_1}(n,n+1)= \ell_{T_1}(n+1,n)=1\\
\ell_{T_1}(n+1,a)=\ell_{T_1}(n,a)>0,\quad\mbox{for every $a\in S\setminus\{n,n+1\}$}\\ 
\ell_{T_1}(a,n+1)=\ell_{T_1}(a,n),\quad\mbox{for every $a\in S\setminus\{n,n+1\}$}
\end{array}
$$
Notice moreover that $n$ is also a deepest leaf  in $T_1$ and therefore, by Lemma \ref{lem:5}, if it is not a deepest leaf in $T_2$, then $d_p^s(T_1,T_2)^p\geq 5$. So, we assume henceforth that $n$ and $n+1$ are deepest leaves in $T_2$.
As in (a), there are several subcases to discuss.

\item[\emph{(b.1)}]  Assume that, in $T_2$, the leaves $n$ and $n+1$ are not
sibling.  In this case,
$$
\ell_{T_2}(n,n+1)=x\geq 1,\quad
\ell_{T_2}(n+1,n)=y\geq 1
$$
and $x>1$ or $y>1$. Since the depths of $n$ and $n+1$ in $T_2$ are the same, it must happen that $x=y$.
Then,
$$
C(n,n+1)+C(n+1,n)=(x-1)^p+(x-1)^p\geq 2.
$$
Let now $a_0\neq n$ a labeled node, other than $n$, that is a descendant of
the parent of $n$ in $T_2$: notice that this parent is an intermediate
node in the path $[n,n+1]_{T_2}\pathgr n$.  Then,
$$
\ell_{T_2}(n,a_0)=1<x=\ell_{T_2}(n+1,a_0),\quad
\ell_{T_2}(a_0,n)<\ell_{T_2}(a_0,n+1)
$$
imply that $a_0$ contributes at least 2 to $d_p^s(T_1,T_2)^p$, and
therefore that $d_p^s(T_1,T_2)^p\geq 4$. Actually, $d_p^s(T_1,T_2)^p\geq 6$, because
any labeled node $b_0\neq n+1$ that is a descendant of
the parent of $n+1$ in $T_2$ will also contribute at least 2 to $d_p^s(T_1,T_2)^p$.

\item[\emph{(b.2)}] Assume that, in $T_2$, the leaves $n$ and $n+1$ are
siblings and their parent is labeled, say with 1.  In this case, by (a) (applied interchanging the roles of $T_1$ and $T_2$ and the roles of $n$ and $1$), we already know that $d_p^s(T_1,T_2)^p\geq 4$.

\item[\emph{(b.3)}] Assume that, in $T_2$, the nodes $n$ and $n+1$ are
sibling leaves and their parent is not labeled.  
 In this case, let $T_1^*,T_2^*\in \nTT_n$ be the trees obtained from $T_1$ and $T_2$, respectively,  by removing the leaves $n$ and $n+1$
together with their pendant arcs, and labeling with $n$ the former
parent of $n$ and $n+1$.  In this way we have that, for every $1\leq
a\neq b\leq n$ and for every $i=1,2$,
$$
\begin{array}{l}
\ell_{T_i^*}(a,b)= \ell_{T_i}(a,b) \quad \mbox{ if $a\neq n$}\\
\ell_{T_i^*}(n,b)= \ell_{T_i}(n,b) -1\quad \mbox{ if $a= n$}
\end{array}
$$
and therefore, $C(a,b)=C_{T_1^*,T_2^*}^p(a,b)$. Then, arguing as in (a.3),
$$
d_p^s(T_1,T_2)^p \geq d_p^s(T_1^*,T_2^*)^p\geq 4.
$$
\end{enumerate}
This finishes the proof by induction. \qed
\end{proof}

\begin{figure}[htb]
\begin{center}
            \begin{tikzpicture}[thick,>=stealth,scale=0.4]
              \draw(0,0) node[btre] (1) {}; \draw (1) node {\scriptsize $i_1$};
              \draw(2,2) node[btre] (2) {}; \draw (2) node {\scriptsize $i_2$};
              \draw(4,4) node[btre] (a) {}; %\etq r
          \draw(6,0) node[btre] (3) {}; \draw (3) node {\scriptsize $i_3$};
                 \draw[->] (a)--(2);
                 \draw[->] (a)--(3);
                 \draw[->] (2)--(1);
           \draw(3,-1.5) node  {$T_1'$};
            \end{tikzpicture}
  \qquad\qquad
             \begin{tikzpicture}[thick,>=stealth,scale=0.4]
              \draw(0,0) node[btre] (1) {}; \draw (1) node {\scriptsize $i_1$};
              \draw(2,2) node[btre] (2) {}; \draw (2) node {\scriptsize $i_3$};
              \draw(4,4) node[btre] (a) {}; %\etq r
          \draw(6,0) node[btre] (3) {}; \draw (3) node {\scriptsize $i_2$};
                 \draw[->] (a)--(2);
                 \draw[->] (a)--(3);
                 \draw[->] (2)--(1);
           \draw(3,-1.5) node  {$T_2'$};
            \end{tikzpicture}
            \\[1ex]
           \begin{tikzpicture}[thick,>=stealth,scale=0.4]
              \draw(0,0) node[btre] (1) {}; \draw (1) node {\scriptsize $i_1$};
              \draw(3,0) node[btre] (2) {}; \draw (2) node {\scriptsize $i_2$};
              \draw(2,2) node[btre] (a) {}; %\etq r
          \draw(4,4) node[btre] (r) {}; %\etq r
        \draw(6,0) node[btre] (3) {}; \draw (3) node {\scriptsize $i_3$};
                     \draw[->] (r)--(a);
                 \draw[->] (a)--(1);
                 \draw[->] (a)--(2);
                 \draw[->] (r)--(3);
           \draw(3,-1.5) node  {$T_1''$};
            \end{tikzpicture}
  \qquad\qquad
          \begin{tikzpicture}[thick,>=stealth,scale=0.4]
              \draw(0,0) node[btre] (1) {}; \draw (1) node {\scriptsize $i_1$};
              \draw(4,0) node[btre] (2) {}; \draw (2) node {\scriptsize $i_2$};
              \draw(2,2) node[btre] (3) {}; \draw (3) node {\scriptsize $i_3$};
          \draw(2,4) node[btre] (r) {}; %\etq r
                     \draw[->] (r)--(3);
                 \draw[->] (3)--(1);
                 \draw[->] (3)--(2);
           \draw(3,-1.5) node  {$T_2''$};
            \end{tikzpicture}
\end{center}
\caption{\label{fig:least1} 
The pairs of basic trees that give rise, when attached to the same place in a tree, to pairs of non-weighted phylogenetic trees at $d_p^s$ distance $\sqrt[p]{4}$.}
\end{figure}
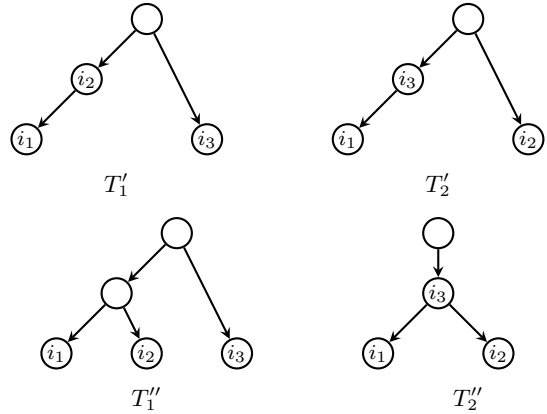

\begin{remark}
Following in detail the arguments developed in the last theorem until their last consequences, it can be proved that, for $n\geq 6$, the pairs of trees $T_1,T_2$ in $\nTT_n$ such that  $d_p^s(T_1,T_2)^p=4$, for every $p\in \NN^+$, 
are exactly those pairs  such that  $d_1(T_1,T_2)=4$, and they have the following form.  Let $i_1,i_2,i_3$ be any three taxa in $S$ and let $T_0$ be any non-weighted rooted tree   with some of its nodes, including all its elementary nodes and all its leaves \emph{except at most one elementary node or one leaf}, labeled in 
 $S\setminus\{i_1,i_2,i_3\}$.
Then,  $T_1$ and $T_2$ are obtained, respectively, by attaching  to $T_0$ at  the same node the `basic'  trees $T_1'$ and $T_2'$ or $T_1''$ and $T_2''$ in Fig.~\ref{fig:least1}. The attachment of one of these  trees at a node $v$ in $T$  is carried  out by identifying the node with the root of the tree, and in such a way that the resulting trees $T_1$ and $T_2$  have all their leaves and elementary nodes labeled.
This implies that if 
 $T$ had some non-labeled leaf or elementary node, this is necessarily the node where the basic trees must be attached, and that (since $T_2''$ has its root elementary), 
 the basic pair $T_1'',T_2''$ cannot be attached to a non-labeled leaf (this would create an elementary node in $T_2$).
 
 For instance, the trees $T$ and $T'$ in Fig.~\ref{fig:dist4} are obtained by attaching the basic trees $T_1'$ and $T_2'$ (with $i_1=1$, $i_2=2$, and $i_3=3$) to the tree with Newick code \texttt{(4,\ldots,n);}.

\end{remark}

\begin{remark}
It can be checked that the pairs of different trees in $\nTT_n$ at least distance for $d_1^s$ have always splitted path lengths matrices with $n-1$ (if $n\leq 5$) or 4 (if $n\geq 5$) entries that differ in only 1. This implies that 
the least non-zero value for $d_\infty^s$ on $\nTT_n$ is always 1, and that the least non-zero value for $d_0^s$  on $\nTT_n$ is again $n-1$ for $n\leq 5$ and 4 for $n\geq 6$.
\end{remark}

Unfortunately, we have not been able to find a formula for the diameter of $\nTT_n$ with respect to any metric $d_p^s$ with $p\in \NN^+$. Actually, and to our knowledge, the diameter of the space of non-weighted binary phylogenetic trees with respect to the nodal metrics $d_1$ and $d_2$ is still not known, either. 
Not knowing a formula for the diameter, we are not able to give an explicit description of the distribution of distances for any $p$, either. In the file \texttt{distributions.pdf} in the Supplementary Material we provide  the distributions of $d_1^s$ and $(d_2^s)^2$ (that is, of $d_2^s$ squared) on $\nTT_n$ for $n=3,4,5,6$, as well as  the distributions  of the values of $d_1^s$ and $(d_2^s)^2$ applied to pairs of trees in TreeBASE sharing $n=2$ to 6 labels.

\section{Conclusions}

Some classical metrics for phylogenetic trees are based on the comparison of the representations of rooted phylogenetic trees as 
vectors of path lengths between pairs of labeled nodes. But these metrics only separate non-weighted binary rooted trees: two more general non-isomorphic rooted phylogenetic trees  can have the same such vectors of path lengths, and therefore be at zero distance for these metrics. In this paper we have overcome this problem by representing  a rooted phylogenetic tree by means of a matrix with rows and columns indexed by taxa and where every entry $(i,j)$ is the distance from the least common ancestor of the pair of nodes labeled with $i$ and $j$ to the node labeled with $i$.
We call these matrices \emph{splitted path lengths matrices}, because they split in two terms the path length between every pair of labeled nodes. These matrices 
define an injective mapping from the space $\TT_n$ of all
$\RRp$-weighted rooted phylogenetic trees with $n$ labeled nodes and possibly nested taxa 
into the set $\mathcal{M}_n(\RR)$ of $n\times n$ real-valued matrices.
Therefore, any norm on $\mathcal{M}_n(\RR)$ applied to the difference of the
splitted path lengths matrices of trees
defines a metric on $\TT_n$. Using the well-known $L^p$ norms on $\mathcal{M}_n(\RR)$, for $p\in \NN\cup\{\infty\}$, we obtain the family of splitted nodal metrics $d_p^s$ on $\TT_n$
$$
d_p^s(T_1,T_2)=\left\{
\begin{array}{ll}
\big|\{(i,j)\mid 1\leq i\neq j\leq n,\  \ell_{T_1}(i,j)\neq \ell_{T_2}(i,j)\}\big| & \mbox{if $p=0$}\\
\sqrt[p]{\sum_{1\leq i\neq j\leq n}^m |\ell_{T_1}(i,j)- \ell_{T_2}(i,j)|^p} & \mbox{if $p\in \NN^+$}\\
\max\{|\ell_{T_1}(i,j)- \ell_{T_2}(i,j)| \mid 1\leq i\neq j\leq n\} & \mbox{if $p=\infty$}
\end{array}
\right.
$$

We have proved several properties for these metrics $d_p^s$  on the subspace $\nTT_n$ of non-weighted rooted phylogenetic trees possibly with nested taxa. For instance, we have established the least distance between any pair of such trees. It remains as an open problem to find the diameter of $\nTT_n$ with respect to these metrics, and the distribution of their values. Actually, these problems also remain open for the classical nodal distances  on
non-weighted binary (rooted as well as  unrooted) trees. These are interesting problems: to know the largest value reached by a metric is necessary to normalize the
metric between 0 and 1, while knowing the distribution of the values allows one to answer the question of whether two trees are more similar than expected by chance \cite{pennyhendy:sz85}. We hope to report on these problems in a near future.

We cannot advocate the use of any splitted nodal metric $d_p^s$ over
the other ones except, perhaps, warning against the use of
$$
\begin{array}{rl}
d_0^s(T_1,T_2) & =\big|\{(i,j)\in S^2\mid \ell_{T_1}(i,j)\neq \ell_{T_2}(i,j)\}\big|\\
d_\infty^s(T_1,T_2) & =\max\{|\ell_{T_1}(i,j)-\ell_{T_2}(i,j)|\mid (i,j)\in S^2\}
\end{array}
$$
because they are too uninformative.
Since the most popular norms on $\RR^m$ are the Manhattan and the
Euclidean, it seems natural to use $d_1^s$ and $d_2^s$, as it
has been the case in the classical, non-weighted binary setting.  Each one
has its advantages.  For instance, the computation of
$d_1^s$ does not involve square roots, and therefore it can be
computed exactly and, if the weights are integer numbers, the resulting
value is an integer number.  Moreover, it is well known that, for
every $p\in \NN^+$,
$$
 \|x\|_p\leq \|x\|_1\quad \mbox{ for every $x\in \RR^m$}
$$
and therefore,
$$
 d_p^s(T_1,T_2)\leq d_1^s(T_1,T_2) \quad\mbox{ for every $T_1,T_2\in \TT_n$}.
$$
On the other hand, the comparison of splitted path lengths matrices by
means of the Euclidean norm enables the use of many geometric
and clustering methods that are not available otherwise.  For instance, the specific properties of the Euclidean norm allowed  Steel and Penny to compute explicitly the mean value of the nodal distance $d_2$ on the class of non-weighted unrooted binary trees \cite{steelpenny:sb93}, while no similar result is known for $d_1$.

As a rule of thumb, we consider suitable to use $d_1^s$ when the trees
are non-weighted (of when they have integer weights), because these
trees can be seen as discrete objects and thus their comparison through a discrete
tool as the Manhattan norm seems appropriate.  When the trees have
arbitrary positive real weights, they should be understood as
belonging to a continuous space \cite{billera.ea:01}, and then the Euclidean norm is more
appropriate.

\section*{Supplementary Material} 

The Supplementary Material referenced in the paper is available
at\\ \texttt{http://bioinfo.uib.es/\~{}recerca/phylotrees/nodal/}.

\section*{Acknowledgements:}
The research described in this paper has been partially supported by
the Spanish DGI projects MTM2006-07773 COMGRIO and
MTM2006-15038-C02-01.

\end{document}